%% file: xip_Y3.tex
\documentclass[prd,amsmath,aps,floats,amssymb, floatfix,superscriptaddress,nofootinbib,twocolumn,preprintnumbers]{revtex4-1}

\usepackage{lineno}

\usepackage{tabularx}
\usepackage{amsmath,amssymb,times}

\usepackage[T1]{fontenc}
\usepackage{xspace}
\usepackage{multirow}

\newcommand{\aap}{A\&A}
\newcommand{\apjs}{ApJS}
\newcommand{\aj}{A.J.}
\newcommand{\mnras}{MNRAS}

\newcommand{\physrep}{Phys. Rep.}

\usepackage{todonotes}
\usepackage{hyperref}
\usepackage{ulem}
\usepackage{bm}

\newcommand{\xip}{\xi_{\rm p}}

\newcommand*\justify{%
  \fontdimen2\font=0.4em
  \fontdimen3\font=0.2em
  \fontdimen4\font=0.1em
  \fontdimen7\font=0.1em
  \hyphenchar\font=`\-
}

\newcommand\code[1]{\texttt{\small\justify #1}}

\newcommand{\ZMEAN}{\code{Z\_MEAN}\xspace}
\newcommand{\ZMC}{\code{Z\_MC}\xspace}

\newcommand{\change}[1]{{\textcolor{black}{#1}}\xspace}

\newcommand{\MpcOh}{ \,  \mathrm{Mpc}  \, h^{-1} }
\newcommand{\hOMpc}{ \,  \mathrm{Mpc}^{-1}  \, h  }
\newcommand{\nn}{ \nonumber }

\newcommand{\beq}{\begin{equation}}
\newcommand{\eeq}{\end{equation}}
\newcommand{\beqa}{\begin{eqnarray}}
\newcommand{\eeqa}{\end{eqnarray}}

\newcommand{\gold}{\code{Y3\,GOLD}\xspace}

\newcommand{\var}[1]{\ensuremath{\texttt{\MakeUppercase{#1}}}\xspace}

\newcommand\be{\begin{equation}}
\newcommand\ee{\end{equation}}
\def\bea{\begin{eqnarray}}
\def\eea{\end{eqnarray}}

\newcommand\T{\rule{0pt}{2.6ex}}       
\newcommand\B{\rule[-1.2ex]{0pt}{0pt}} 

\begin{document}
\title{ Dark Energy Survey Year 3 Results: Measurement of the Baryon Acoustic Oscillations \\
  with   Three-dimensional Clustering }

\input{author_list.tex}


\noaffiliation

\date{\today}
\label{firstpage}
\begin{abstract}

  The three-dimensional correlation function offers an effective way to summarize the correlation of the large-scale structure even for imaging galaxy surveys. We have applied the projected three-dimensional correlation function, $\xip$  to measure the Baryonic Acoustic Oscillations (BAO) scale on the first-three years Dark Energy Survey data.  The sample consists of about 7 million galaxies in the redshift range  $ 0.6 < z_{\rm p } < 1.1 $ over a footprint of $4108 \, \mathrm{deg}^2 $.  Our theory modeling includes  the impact of realistic true redshift distributions beyond Gaussian photo-$z$ approximation.  $ \xip $ is obtained by projecting the three-dimensional correlation to the transverse direction.  To increase the signal-to-noise of the measurements, we have considered a Gaussian stacking window function in place of the commonly used top-hat.  $\xip$ is sensitive to  $ D_{\rm M}(z_{\rm eff} )  / r_{\rm s} $, the ratio between the comoving angular diameter distance and the sound horizon.  Using the full sample, $ D_{\rm M}(z_{\rm eff} )  / r_{\rm s} $  is constrained to be $ 19.00 \pm 0.67 $ (top-hat) and  $ 19.15 \pm 0.58 $ (Gaussian) at  $z_{\rm eff} = 0.835$.  The constraint is weaker than the angular correlation $w$ constraint ($18.84 \pm 0.50$), and we trace this to the fact that  the  BAO signals are heterogeneous across redshift.  While $\xip $ responds to the heterogeneous signals by enlarging the error bar,  $w$ can still give a  tight bound on  $D_{\rm M}  / r_{\rm s} $ in this case.  When a homogeneous BAO-signal sub-sample in the range  $ 0.7 < z_{\rm p } < 1.0 $ ($z_{\rm eff} = 0.845$) is considered,  $\xip $ yields  $ 19.80 \pm 0.67  $ (top-hat) and $ 19.84 \pm 0.53 $ (Gaussian). The latter is mildly stronger than the $w$ constraint ($19.86 \pm 0.55 $). We find that the $\xip $ results are  more sensitive to photo-$z$ errors than $w$ because $\xip$ keeps the three-dimensional clustering information causing it to be more prone to photo-$z$ noise.  The Gaussian window gives more robust results than the top-hat as the former is designed to suppress the low signal modes.   $\xip $ and the angular statistics such as $w$ have their own pros and cons, and they serve an important crosscheck with each other.

\end{abstract}


\preprint{DES-2022-0707}
\preprint{FERMILAB-PUB-22-728}

\maketitle

\section{ Introduction }

The Baryonic Acoustic Oscillations (BAO)  \cite{PeeblesYu1970,SunyaevZeldovich1970}  has been recognized as one of the most important probes in cosmology. It is the primordial acoustic features imprinted in the distribution of the large-scale structure. In the early universe, photons tightly couple with the baryons (electrons and protons) to form a plasma and acoustic oscillations are excited. The sound waves propagate until the recombination time, after which the plasma ceases to exist  and the acoustic waves are stalled. The acoustic patterns are preserved in the large-scale structure, and the characteristic scale encoded corresponds to the sound horizon at the drag epoch, which is about 150 Mpc in standard cosmology. Since the physics for BAO formation is linear and well-understood (e.g.~\cite{BondEfstathiou1984,BondEfstathiou1987,HuSugiyama1996,HuSugiyamaSilk1997,Dodelson_2003}), the sound horizon scale can be computed to high precision and the BAO is widely regarded as a standard ruler \cite{WeinbergMortonson_etal2013,Aubourg:2014yra}.  Ever since its clear detection in SDSS  \cite{Eisenstein_etal2005}  and 2dFGS \cite{Cole_etal2005}, the BAO measurements have been repeated using numerous spectroscopic datasets at different effective redshifts \cite{Gaztanaga:2008xz, Percival_etal2010, Beutler_etal2011, Blake2012_WiggleZ, Anderson_BOSS2012, Kazin_etal2014, Ross_etal2015, Alam_etal2017, eBOSS:2020yzd}.

Imaging surveys are another type of major galaxy surveys, in which the redshift of galaxies, photo-$z$, is inferred by means of a few broadband filters. There are a number of ongoing and future large-scale photometric surveys including the Kilo-Degree Survey (KiDS) \footnote{http://kids.strw.leidenuniv.nl}, Dark Energy Survey (DES) \footnote{https://www.darkenergysurvey.org},  Hyper Suprime-Cam (HSC) \footnote{https://www.naoj.org/Projects/HSC}, Rubin Observatory's Legacy Survey of Space and Time (LSST) \footnote{https://www.lsst.org}, Euclid \footnote{https://www.euclid-ec.org}, and the Chinese Survey Space Telescope (CSST) \footnote{http://www.nao.cas.cn/csst}.  While the precision of photo-$z$ is limited, for instance the photo-$z$ accuracy  for  the bright red galaxies in DES is about $\sigma \sim 0.03(1+z)$ \cite{Crocce_etal2019,DES:2021jns}, the photometric surveys can collect a large volume of data with deep magnitude efficiently.

Photometric data suffer from photo-$z$ smearing in the radial direction, but the information in the transverse direction remains intact. Thanks to their large data volume and deep magnitude, competitive BAO measurements can be obtained  \cite{SeoEisenstein_2003,BlakeBridle_2005,Amendola_etal2005,Benitez:2008fs,Zhan:2008jh, Chaves-Montero:2016nmw}.  This has been demonstrated using photometric data by various groups: SDSS \cite{Padmanabhan_etal2007,EstradaSefusattiFrieman2009, Hutsi2010, Seo_etal2012,Carnero_etal2012, deSimoni_etal2013}, DES [Y1 \cite{Abbott:2017wcz} and Y3 \cite{DES:2021esc}, hereafter DES Y3],  and DECaLS \cite{Sridhar_etal2020}.  In particular, DES Y3 measured the BAO at the effective redshift of 0.835 and constrained the comoving angular diameter distance divided by the sound horizon scale to be $ 18.92\pm 0.51 $. This constraint is tighter than the corresponding result from eBOSS ELG sample at a similar redshift \cite{Tamone_etal2020,deMattia_etal2021} by roughly a factor of two.  This example highlights that the photometric galaxy clustering analysis indeed can deliver strong cosmological constraints.

The DES Y3 BAO measurements are performed in two angular statistics: the angular correlation function in configuration space and the angular power spectrum in harmonic space, and their results are well consistent with each other. Overall, the treatments of these angular statistics are similar and highly correlated.  In these angular tomographic analyses, the data in the whole redshift range [0.6,1.1] are divided into five tomographic bins of equal width.  Only the auto-correlation function is considered, but not the cross correlation. This is mainly because the BAO information in the cross correlation is still limited in current survey size \cite{Chan:2018gtc}. Besides, for the tomographic analysis, inclusion of the cross correlation would increase the size of the data vector substantially.   Alternatively, the photometric data can be analyzed using the three-dimensional correlation  akin to the spectroscopic analysis \cite{Ross:2017emc}. In this method, we use the three-dimensional position of the galaxies deduced from photo-$z$ to compute the spatial correlation, which is then projected to the transverse direction.  We shall abbreviate this statistic as $\xip$.  The initial modeling proposed in \cite{Ross:2017emc} was limited to Gaussian photo-$z$ approximation.  Nonetheless, it had been applied to photometric survey data to get promising results \cite{Abbott:2017wcz,Sridhar_etal2020}.  To avoid the possibility of introducing bias due to Gaussian photo-$z$ approximation,  however, the $ \xip$ method was not adopted in the DES Y3 key BAO analysis.  Recently, the theory for $\xip$ has been further developed \cite{Chan_etal2021}. Among other things, the modeling is generalized to incorporate arbitrary photo-$z$ uncertainties. The advantage of the $ \xip$ statistic is that it can effectively compress the information into a data vector appreciably smaller in size. The cross correlation information is included automatically.  However, due to photo-$z$ mixing, the  $ \xip$ covariance has large off-diagonal elements, which  cause some troubles for the analysis.  Nonetheless, it can be circumvented for the BAO analysis. Based on mock tests, the improved $  \xip$ method was demonstrated to give a statistically  mildly stronger measurement than the angular correlation analysis result \cite{Chan_etal2021}, but an application to the actual data is still lacking. Furthermore, the DES Y3 BAO analysis yielded an interesting  $ 2 \sigma $ deviation from the Planck result \cite{Planck2020}, it is imperative to crosscheck it using an alternative statistic as they have different sensitivities to potential systematics. Thus it is the goal of this paper to apply the  $ \xip$  statistic to measure BAO on the DES Y3 data.

This paper is organized as follows. We present the properties of the galaxy sample used in this work and discuss its photo-$z$ measurement and calibration in Sec.~\ref{sec:sample_properties}.  In Sec.~\ref{sec:analysis_pipeline}, we first review the computation of the $\xip$ template and the covariance, and then describe the procedures for parameter inference. We present some mock test results in Sec.~\ref{sec:mock_tests}. Our main results are in Sec.~\ref{sec:results}, where we show the measurement of the BAO and the robustness tests conducted to check the soundness of the results. We pay particular attention to contrast the $ \xip $ results against those from the angular correlation function. Sec.~\ref{sec:conslusions} is devoted to the conclusions. The pre-unblinding test results are shown in Appendix \ref{sec:PreunblindingTest}.   In Appendix \ref{appendix:errorbars}, we test the impact of heterogeneity in the BAO signals on the error bar through mocks.   The default cosmology for the data analysis is a flat $\Lambda $CDM in the Planck cosmology  \cite{Planck2020} with  $\Omega_{\rm m} = 0.31$, $h = 0.676$, $n_s=0.97$, and $\sigma_8=0.83$ (abbreviated as Planck hereafter). Because the mock catalogs were created in the MICE cosmology  \citep{Fosalba_etal2015,Crocce_etal2015}, which is a flat $ \Lambda$CDM with $\Omega_{\rm m} = 0.25$,  $\Omega_{\Lambda} = 0.75$, $h=0.7$, and $\sigma_8 = 0.8$  (denoted as MICE), we also consider adopting the MICE cosmology as an alternative.

\begin{figure}[!htb]
\centering
\includegraphics[width=\linewidth]{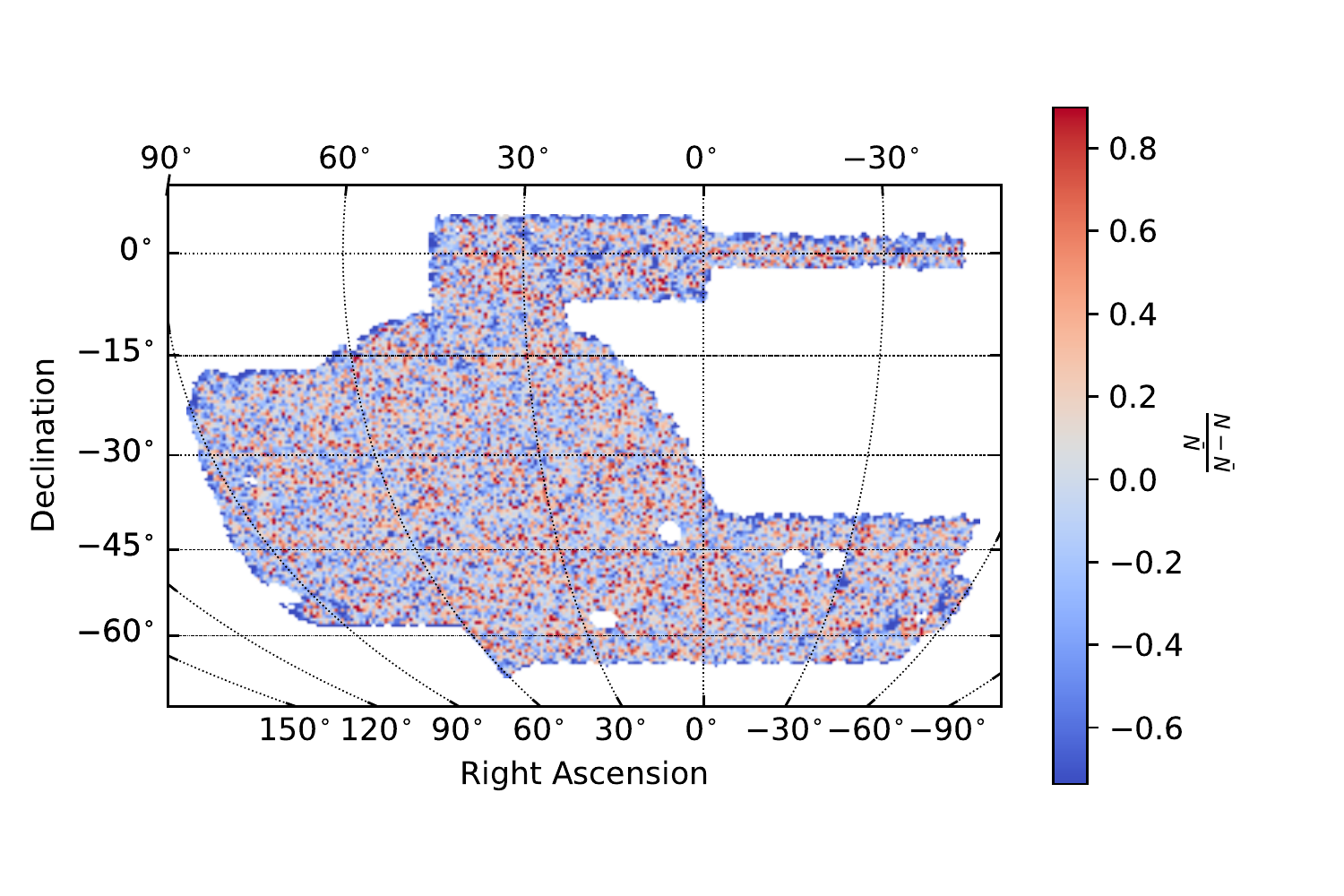}
\caption{   The footprint of the BAO sample used in this work.  Shown here is the galaxy density contrast at the pixel scale (Healpix resolution of 512).  The full DES Y3 footprint spans nearly 5000 deg$^2$, but after various cuts, the footprint for the BAO sample is reduced to a total area of 4108 deg$^2$.       }
  \label{fig:baosample_dist}
\end{figure}

\section{Sample data properties } 
\label{sec:sample_properties}

In this section, we first introduce the galaxy sample used for BAO measurement and then move to describe the photo-$z$ estimation and its calibration for this sample.

\subsection{Galaxy sample }

In this subsection, we describe the DES Y3 galaxy sample, called the BAO sample hereafter,  on which the BAO scale is measured.  Through the angular correlation function and the angular power spectrum analyses, it was previously used to obtain the most precise measurement of the BAO from  photometric data \cite{DES:2021esc}.  Here we outline the essential information about the sample, and refer the readers to  \cite{DES:2021jns} for more details.

The BAO sample was built from the DES first three-year  (Y3) data, which were observed by the Dark Energy Camera (DECam) \cite{Flaugher_etal2015} at the Blanco 4 m telescope at the Cerro Tololo Inter-American Observatory in Chile. The raw data cover about 5000 $\mathrm{deg}^2$  in the southern sky and include observations in  five photometric bandpasses $grizY$.  The data were made available to the public in DR1 \cite{DES_DR1}.  After further processing and improvements, the \gold sample \cite{y3-gold} suitable for cosmological analyses was assembled.  This sample comprises of 390 million galaxies with  $i$-band limiting magnitude up to 23 (AB, 10 $\sigma$ level).

Red galaxies tend to be old galaxies that passively evolve with time. They are often hosted in massive halos with significant galaxy bias.  Thus they furnish a good tracer of the large-scale structure.  Furthermore, because there are more features in their SED, their photo-$z$ quality tends to be better than the ones for blue galaxies.  The BAO sample is a red galaxy sample constructed out of the \gold sample by applying  color cuts following the Year 1 sample definition \cite{Crocce_etal2019}.  The precise cuts in magnitude and photo-$z$ $z_{\rm p}$  are given by \cite{DES:2021jns}
\begin{align}
(i_\var{SOF}-z_\var{SOF}) + & 2.0 (r_\var{SOF}-i_\var{SOF})  > 1.7  ,  \label{eq:colorcut}  \\ 
 i_\var{SOF}  <  &19 + 3.0 \, z_{\rm p}     ,  \label{eq:magcut} \\
0.6< & z_{\rm p}  <1.1  , \label{eq:zrange}
\end{align} 
where  \var{SOF} signifies the Single Object Fitting method used to derive the magnitude. The cuts take into account the trade-off between the number density and the photo-$z$ quality. In addition, a bright magnitude cut $  i_\var{SOF}>17.5 $ is imposed to get rid of bright contaminants such as binary stars, and  objects that are deemed  suspicious or problematic are also removed. The star-galaxy separation is performed with the \var{EXTENDED\_CLASS\_MASH\_SOF} flag in the \gold catalog, and the star contamination on the BAO sample is estimated to be under a few percent \cite{DES:2021jns}.

\begin{figure}[!tb]
\centering
\includegraphics[width=\linewidth]{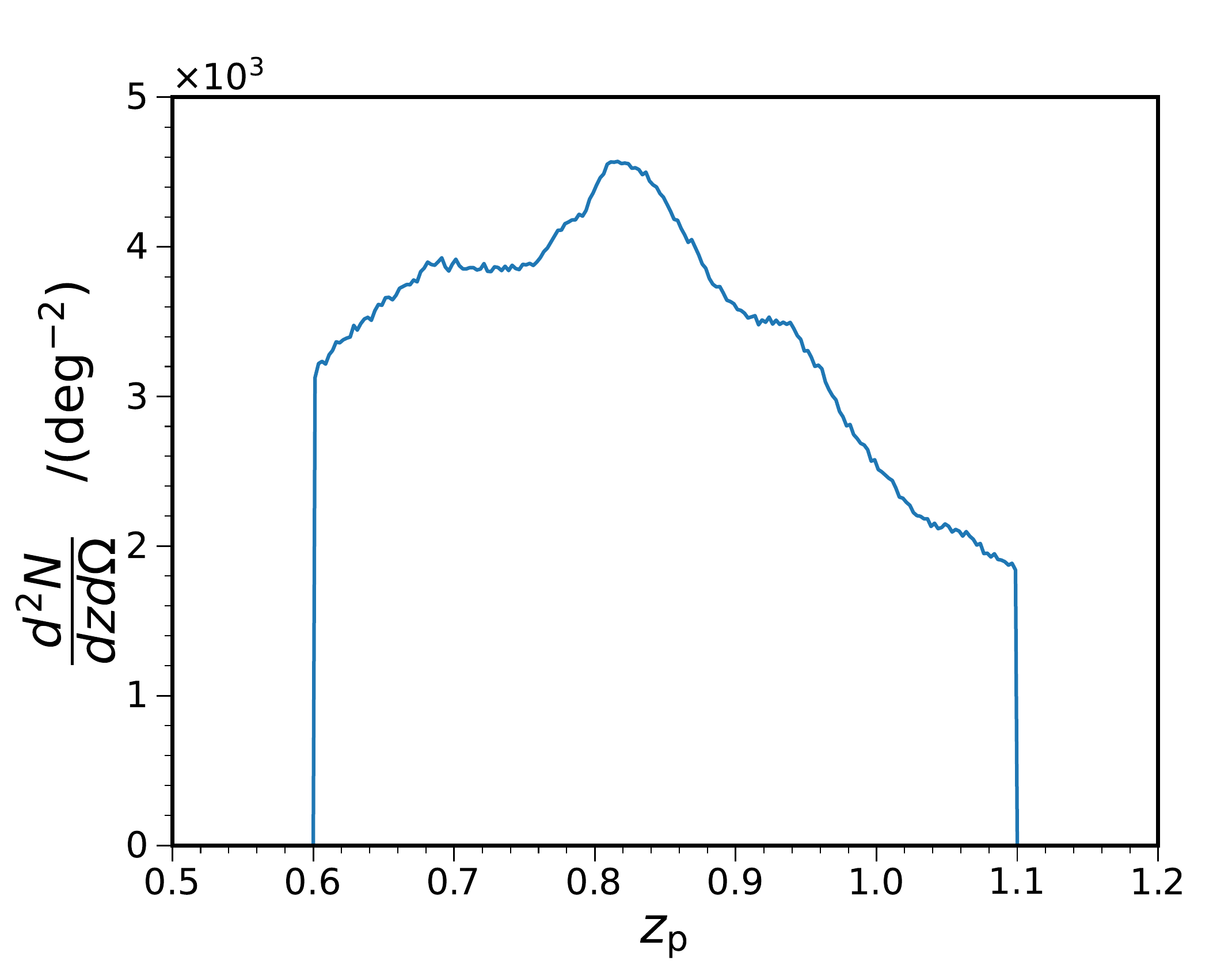}
\caption{ The number of galaxies per unit redshift per unit squared degree for the BAO sample as a function of the photo-$z$ best fit. }
  \label{fig:dN_dzdOmega}
\end{figure}

The resultant BAO sample consists of 7.03 million galaxies in the redshift range [0.6,1.1]  with $i$-band limiting magnitude $i<22.3$.  The BAO sample footprint is shown in Fig.~\ref{fig:baosample_dist}. In Healpix resolution of $ N_{\rm side} = 4096 $, each pixel is covered at least once in  $griz$ with coverage greater than 80\%. After the foreground and other removals, the effective area of the survey mask totals 4108 $\mathrm{deg}^2$.  In Fig.~\ref{fig:dN_dzdOmega}, we show the number of galaxies per unit redshift per unit squared degree for the BAO sample. The effective redshift of the sample is 0.835.

Because the observations are taken over a long period of time and in large spatial locations, the data are unavoidably affected by the observational conditions (survey properties). These effects may give rise to spurious signals if not corrected for. From over 100 correlated survey property maps available for \gold, using a principle component analysis technique, 26  systematic property principle component maps including the depth, air mass, stellar density, and extinction, are extracted. These maps are used to calibrate the systematic correction weights. The systematic weights are applied to the galaxies iteratively until there is no appreciable dependence of the galaxy density on the survey properties. See \cite{DES:2021jns} for more details on systematic corrections of the BAO sample.

\subsection{Photo-$z$ and its calibration}

\begin{figure}[!htb]
\centering
\includegraphics[width=\linewidth]{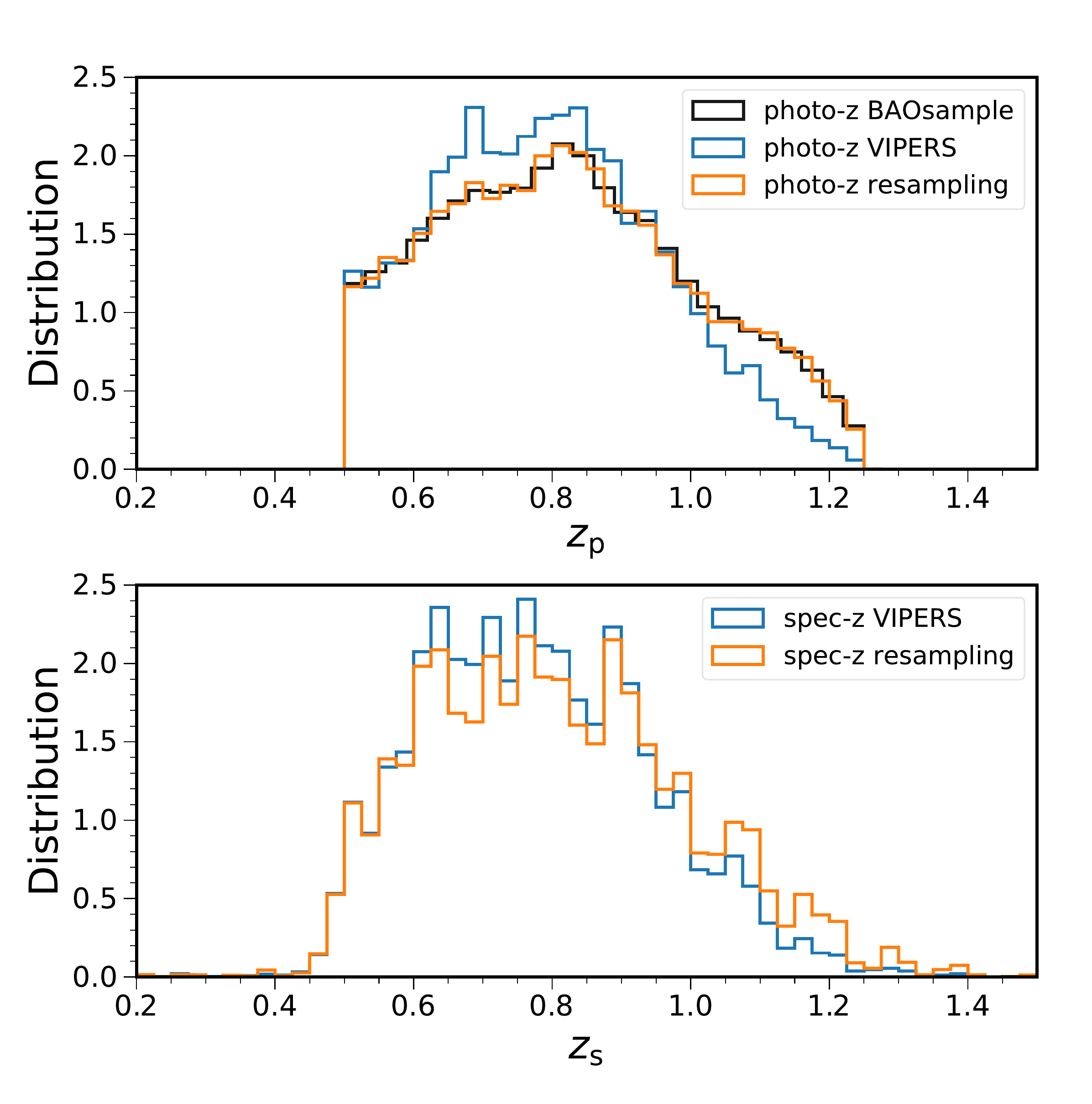}
\caption{  Upper panel: The photo-$z$ distribution of the BAO sample (black), the VIPERS sample (blue), and the resampled  VIPERS (orange). While there are some differences between the BAO sample and the VIPERS, the BAO sample agrees with  the resampled  VIPERS well by construction. Lower panel: The spec-$z$ distribution of the  VIPERS (blue) is in good agreement with the resampled VIPERS result (orange), which is a more accurate representation of the spec-$z$ distribution of the BAO sample. }
\label{fig:zph_zpec_vipers_resampling}
\end{figure}

The photo-$z$ of the galaxies is computed by the Directional Neighborhood Fitting (DNF) algorithm \cite{DeVicente:2015kyp}  based on the data in $griz$ bands.  DNF is a training method, and it predicts the best-fit photo-$z$ estimate (\ZMEAN) by performing a nearest-neighbors fit to the hyper-plane in color-magnitude space of the training sample.   We use \ZMEAN as the primary redshift estimate.
 Moreover, DNF outputs a second proxy for the redshift estimate (\ZMC), which is the nearest neighbor redshift in its training set.  The ensemble of \ZMC values serves as a proxy for the total $n(z)$  of a given selection.
A large spectroscopic dataset including up to about  $2.2 \times 10^5 $ galaxies from 24 different spectroscopic surveys available in 2018 are used for training, with the SDSS DR14 \cite{sdssdr14} and the OzDES program \cite{ozdes:2017} as the noted examples.  

To calibrate the photo-$z$ accuracy and to measure the true redshift distribution $\phi$ (see Eq.~\eqref{eq:conditional_true_z} below), we use the spectroscopic data from the VIPERS survey \cite{VIPERS}. An area of 16.32 $ \mathrm{deg}^2 $  overlapping with the DES footprint is used, in which there are  12088 galaxies matching to the DES BAO sample \footnote{There are 74591 VIPERS galaxies before matching to the BAO sample. }. From now on, VIPERS sample always refers to the sample matching to the DES BAO sample.  By binning this galaxy sample based on their photo-$z$ values, the true redshift distribution for the photo-$z$ bin can be estimated.  Notice that the distribution is obtained by counting galaxies, the resultant distribution is implicitly weighted by the underlying spectroscopic number density $ n(z) $, and hence it indeed furnishes an estimate of $\phi$.


We verify the consistency of these procedures by resampling the VIPERS galaxy distribution. In Fig.~\ref{fig:zph_zpec_vipers_resampling}, we show the normalized distribution of the BAO sample galaxies in the photo-$z$ range of [0.5, 1.25] together with the photo-$z$ distribution of the VIPERS sample. We find that there are indeed some differences between them.  Note that this redshift range is wider than that of the final BAO sample, [0.6,1.1].  The VIPERS sample is built by matching the angular position of galaxies from the BAO sample and the original VIPERS sample, and so the matched sample includes only the overlapping part of their distributions.   The VIPERS galaxies can be thought of as a resampling of the  BAO sample with \textit{some distribution}, and its photo-$z$ distribution does not necessarily coincide with that of the BAO sample.

By resampling the VIPERS galaxies using the photo-$z$ distribution of the BAO sample, we can ensure that they match by construction.   Resampling is a weighting of the original data, and our method is similar to the bootstrap resampling with replacement \cite{Efron_resampling}.  Formally we can express this as     
\beq
n_{\rm DES}( z_{\rm s} ) \approx \int d z_{\rm p}  \frac{  n_{\rm VIPERS} ( z_{\rm s} )  n_{\rm DES} (z_{\rm p} ) }{  n_{\rm VIPERS} (z_{\rm p} ) }  g_{\rm VIPERS}(z_{\rm p} | z_{\rm s} ) ,
\eeq
where $ g_{\rm VIPERS} (z_{\rm p} | z_{\rm s} ) $ is the photo-$z$ probability density conditional on the spectroscopic redshift $z_{\rm s}$.
The spectroscopic redshift distribution of the resampled VIPERS galaxies is also shown in Fig.~\ref{fig:zph_zpec_vipers_resampling}. Except for small impacts in the redshift range [0.6,0.8] and [1.0,1.2], overall the resampled  spec-$z$ galaxy distribution is in good agreement with the original VIPERS galaxy spec-$z$ distribution. We note that the resampled VIPERS spec-$z$ distribution should be a more accurate representation of the spec-$z$ distribution of the BAO sample.

\begin{figure*}[!htb]
\centering
\includegraphics[width=\linewidth]{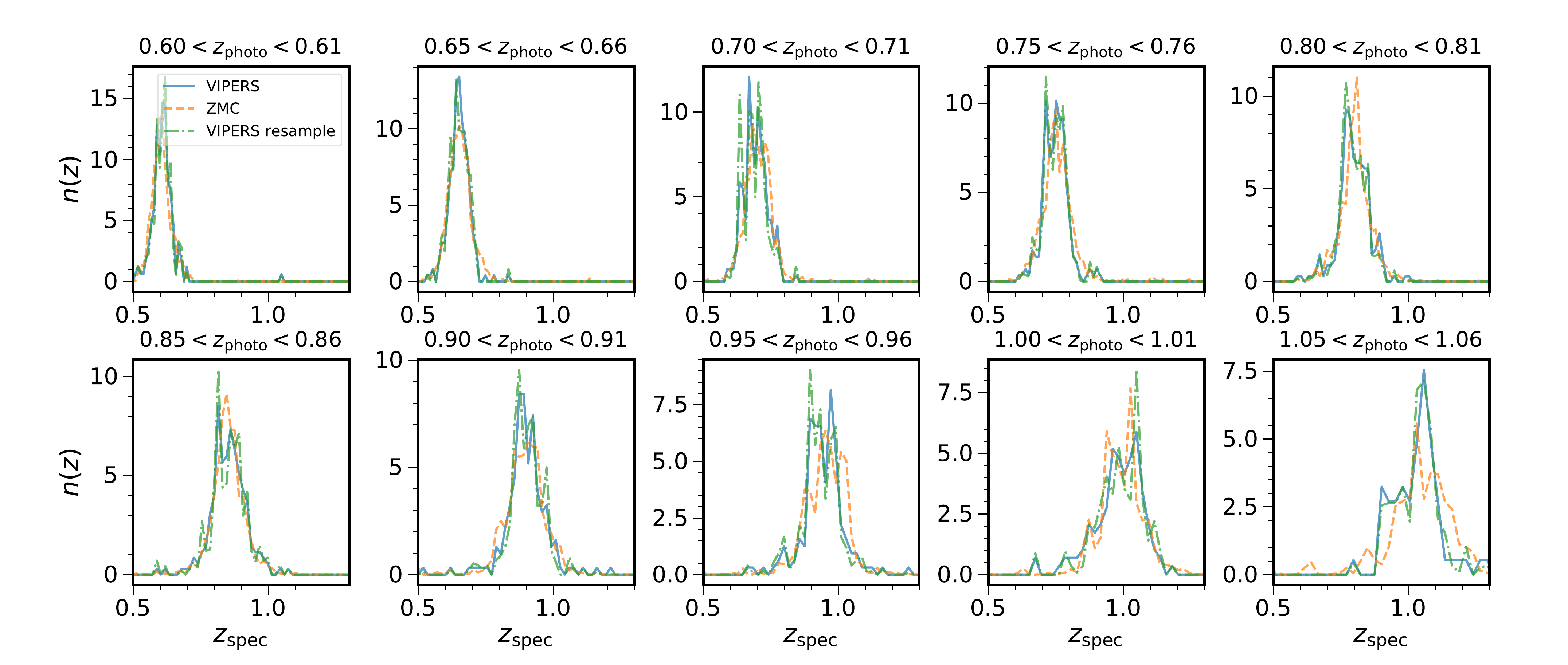}
\caption{ The estimate of the true redshift distribution for the BAO sample calibrated with the VIPERS sample (blue, solid),  the  \ZMC (orange, dashed), and the resampled VIPERS (green, dotted-dashed) for photo-$z$ bins of width $ \Delta z_{\rm p} =  0.01$. Only a subset of the photo-$z$ bin results are shown.   }
  \label{fig:dndz_VIPERS_ZMC_resample_50bins}
\end{figure*}

In Fig.~\ref{fig:dndz_VIPERS_ZMC_resample_50bins}, we show the true redshift distribution estimated from the VIPERS sample, and \ZMC  and the resampled VIPERS in fine photo-$z$ bin of width $ \Delta z_{\rm p} = 0.01 $.  Ref.~\cite{DES:2021jns} found that the $n(z)$ estimate from VIPERS  is more  accurate than that from  \ZMC, and hence the  VIPERS estimation is used as the fiducial choice.  Here we find that the resampled VIPERS results are  in good agreement with the VIPERS ones,  and so from now on, we will only consider the distribution estimated from the VIPERS.  These distributions will be used to compute the theory template and the theory Gaussian covariance.

The resampling method is more useful when there are larger differences between the original distribution and the resampled one. This can happen in e.g.~DES Y6 because the photometric sample is expected to be deeper in magnitude and higher in redshift and it is challenging for the reference spectroscopic sample to match, especially in the high redshift end.

\section{ Analysis pipeline  }
\label{sec:analysis_pipeline}

In this section, we first review how to compute the theory template and the Gaussian covariance for the $\xip$ statistics. We then discuss the fitting method used to extract the BAO scale from the data.

\subsection{ $\xip$ theory template and covariance }
\label{sec:theory}

Here we review the method for computing the $ \xip$ template and its covariance in \cite{Chan_etal2021}.  The basic idea is to map the general cross angular correlation function $ w_{ij}(\theta) $ to  $ \xip$. This takes advantage of the fact that the machinery for the angular correlation function has been well-developed. In particular, this method can easily include general photo-$z$ distributions.

As in the conventional angular tomographic analysis, the whole redshift range is divided into a number of redshift bins, but the number is much larger to ensure that the conditional true redshift distribution  $\phi$ and the bias parameter $b$ approach the intrinsic ones. The conditional true redshift distribution  $\phi(z|z_{\rm p}) $ is central to the clustering analysis of the photo-$z$ data  and is given by
\beq
\label{eq:conditional_true_z}
\phi( z|z_{\rm p} ) = f ( z|z_{\rm p} )  \frac{ \bar{n}(z) }{ \bar{n}_{\rm p} (z_{\rm p}  ) } ,
\eeq
where $ \bar{n} $ and  $ \bar{n}_{\rm p} $  are the mean number density in spectroscopic and photometric redshift space respectively, and $  f ( z|z_{\rm p} )  $ is the  conditional probability density for the  true redshift $z$ given the photo-$z$ being  $ z_{\rm p} $.  As mentioned, we estimate $\phi$  with the help of the spectroscopic sample from VIPERS survey \cite{VIPERS}.   The linear bias parameters  $b$ for the tomographic bins are measured using the angular auto correlation.

With these ingredients, we can compute the general cross angular correlation function $ w_{ij}( \theta_k) $ between the photo-$z$ bin $i$ ($z_{\rm p}$) and  $j$ ($z_{\rm p}'$)  \citep{Hamilton1992,ColeFisherWeinberg1994,CrocceCabreGazta_2011}
\begin{align}
  w_{ij}( \theta ) & = \sum_{\ell=0,2,4} i^\ell   \int dz \phi(z|z_{\rm p})   \int dz' \phi(z'|z_{\rm p}')   \nn \\
  &  \times \mathcal{L}_\ell( \hat{\bm{s}} \cdot \hat{\bm{e}} ) 
  \int \frac{d k k^2  }{ 2 \pi^2  } j_\ell(ks) P_\ell(k,z,z') ,
\end{align}
where  $\mathcal{L}_\ell $ and $ j_\ell $ are the Legendre polynomial and the spherical Bessel function, and $ \hat{\bm{s}} \cdot \hat{\bm{e}} $ is the dot product between the direction of the separation vector of the pair, $ \bm{s} $ and   the line-of-sight direction $\hat{\bm{e}}$. For convenience, we generally use $ z_{\rm p}$  to refer to either the photo-$z$ of an individual galaxy or the photo-$z$ bin if no confusion arises.

The power spectrum multipole $  P_\ell$ is related to the power spectrum $P$ by
\beq
P_\ell(k,z,z')  = \frac{ 2 \ell + 1 }{ 2 }  \int_{-1}^1  d \mu P(k,\mu, z,z') \mathcal{L}_\ell (\mu) . 
\eeq
As in \cite{DES:2021esc}, we use the linear redshift-space power spectrum \cite{Kaiser87} with anisotropic BAO damping:
\begin{align}
  \label{eq:Pk_anisotropic_damp}  
  P( k,\mu, z, z' ) & = [ b + f  \mu^2 ]  [ b' + f'  \mu^2 ]  D(z) D(z')  \nn \\
 &  \times [ ( P_{\rm lin} - P_{\rm nw} )e^{ -k^2 \Sigma_{\rm tot}^2(\mu) }  + P_{\rm nw} ],  
\end{align}
where  $D$ is the linear growth factor and $ f = d \ln D / d \ln a $, and  $ P_{\rm lin } $ and  $  P_{\rm nw} $ denote the linear power spectrum and the smooth power spectrum without BAO information.  The BAO feature is smoothed anisotropically by the damping factor $ \Sigma_{\rm tot}^2(\mu) $, which is computed analytically using the IR resummation \cite{Senatore:2014via,Baldauf:2015xfa,Blas:2016sfa,Ivanov:2018gjr}. Here we follow \cite{Ivanov:2018gjr} to compute  $ \Sigma_{\rm tot}^2(\mu) $ as
\beq
\label{eq:Sigma_tot} 
\Sigma^2 (\mu) = \mu^2 \Sigma_{\parallel}^2 + (1-\mu^2)\Sigma_{\perp}^2+f\mu^2(\mu^2-1)\delta\Sigma^2 ,
\eeq
where $\Sigma_{\parallel}=(1+f) \Sigma$ and $\Sigma_{\perp}=\Sigma $, and
\begin{align}
 \Sigma^2  & = \frac{1}{6\pi^2}\int_0^{ k_{\rm s} } dq \, P_{\rm nw}(q) \left[ 1- j_0(q L ) + 2 j_2(q L)\right],    \\
 \delta \Sigma^2 & = \frac{1}{2\pi^2}\int_0^{ k_{\rm s} } dq \,P_{\rm nw}(q) j_2(q L) , 
\end{align}
where $L$ is the correlation length of BAO. Taking  $L = 110 \MpcOh$ and  $ k_{\rm s}  = 0.2 \hOMpc$, for the MICE cosmology we obtain $\Sigma = 5.80  \MpcOh$ and $\delta \Sigma = 3.18 \MpcOh$ while for the  Planck cosmology  we find $\Sigma = 5.30\MpcOh$ and $\delta \Sigma=2.81\MpcOh$.

After assuming a fiducial cosmology, angles and redshifts can be converted to the separation distance $s$ and its dot product with the line of sight direction $\mu$.   To mimick the measurement of $\xip$ from the data, we loop over  $ w_{ij}( \theta_k) $ and bin it into $s$ and $\mu$.  That is, $\xip$ can be expressed as a weighted mean of  $ w_{ij}( \theta_k) $
\begin{align}
  \label{eq:xi_w_weighting} 
\xi_{\rm p} (s,\mu) = \frac{\sum_{ ijk } f_{ ijk } w_{ij} ( \theta_k  )   }{\sum_{ ijk } f_{ ijk } } , 
\end{align}
where $f_{ijk}$ denotes the weight for all the cross bin pairs $ w_{ij}(\theta_k)$  falling into the $s$  and $\mu $ bins.  This approach enables us to compute $ \xip$ with general photo-$z$ distribution, and it should work as long as the bin size is small compared to the width of $n(z)$ and the intrinsic clustering scale.

Even though we consider the three-dimensional correlation, it only effectively probes the transverse BAO scale, and so the BAO feature lines up at the transverse scale $s_\perp \equiv  s \sqrt{1 - \mu^2 }$ rather than $s$  \cite{Ross:2017emc}. This can be understood as an interplay between the true redshift distribution due to the photo-$z$ uncertainty and the Jacobian of the coordinate transformation \cite{Chan_etal2021}. For  $\sigma_z \gtrsim 0.02 ( 1 +z )$, while the true redshift distribution peaks at the true BAO scales smoothly, because the Jacobian diverges at the transverse scale, the integral is dominated by correlation function at $s_\perp$.  Unfortunately, this also means that $ \xip$ cannot be used to directly probe the Hubble parameter.

To increase the signal-to-noise of the measurement, we stack the measurement of $ \xi_{\rm p} (s ,\mu) $ with different $ \mu$ together:
\beq
\xip( s_\perp ) = \frac{ \sum_i \xip( s, \mu_i ) W(\mu_i)  }{ \sum_i W(\mu_i)  },
\eeq
where $W(\mu)$ is the stacking weight.  Because we effectively project  $ \xi_{\rm p} (s ,\mu) $ along the line-of-sight, we also refer to $\xi_{\rm p} ( s_\perp )  $ as the projected three-dimensional correlation function.  Following \cite{ Ross:2017emc}, previous analyses, including \cite{Chan_etal2021}, consider a top-hat window  $W_{\rm TH}$ 
\beq
\label{eq:TH_stacking}
W_{\rm TH}( \mu, \mu_{\rm max } )  =
    \begin{cases}
        1 &  \textrm{ if  $ \mu < \mu_{\rm max }  $, } \\
        0 &   \text{otherwise,  } \\
    \end{cases}
\eeq
with $  \mu_{\rm max }=0.8 $. Here we assume that  $ \mu \ge 0$. However, the strength of the signal decreases as $\mu$ increases because the effective true redshift distribution becomes wider and less sharply corresponds to the transverse scale (see Fig.~2 in \cite{Chan_etal2021}). Stacking the pairs of different $\mu$ with equal weight is sub-optimal. This motivates us to consider a cut-off Gaussian  $W_{\rm G}$  defined as 
\beq
\label{eq:cutoffGaussian_stacking}
W_{\rm G}( \mu, \sigma_{\mu} )  =
    \begin{cases}
      \exp \big({ - \frac{ \mu^2 }{ 2 \sigma_\mu^2 }   } \big)   &  \textrm{ if  $ \mu < \mu_{\rm max }  $} , \\
        0 &   \text{otherwise . } \\
    \end{cases}
\eeq
This stacking window gives more weight to the small-$\mu$ pairs.

The covariance of $ \xi_{\rm p}( s_\perp ) $ is 
\begin{align}
   &  \quad \quad \mathrm{Cov}[ \xi_{\rm p}( s_\perp ), \xi_{\rm p}( s'_{\perp} ) ]  \nn \\
   = &  \frac{ \sum_i \sum_j    W(\mu_i)  W(\mu_j)   \mathrm{Cov}(  \xi_{\rm p}( s, \mu_i ),  \xi_{\rm p}( s', \mu_j )  )         }{  \sum_i W(\mu_i)  \sum_j W(\mu_j)    }  .
\end{align}
Since the method allows us to map $w$ to $ \xip$, the same mapping also provides a means to derive the covariance for $ \xip ( s, \mu) $ in terms of the covariance of $ w_{ij}(\theta)$: 
\begin{align}
   &  \quad \quad \mathrm{Cov}[ \xi_{\rm p}(s, \mu), \xi_{\rm p}(s', \mu') ]  \nn \\
   = & \frac{ \sum_{ ijk } \sum_{ lmn }  f_{ijk} f_{lmn} \mathrm{Cov}( w_{ij} ( \theta_k ) , w_{lm} ( \theta'_n )    )     }{ \sum_{ijk } f_{ijk}  \sum_{lmn } f_{lmn}  }. 
\end{align}
The general Gaussian covariance for the angular correlation function can be written in terms of the  angular power spectrum  $C_\ell $ as \cite{CrocceCabreGazta_2011}
\begin{align}
  \label{eq:cov_mat_crossz}
& \mathrm{Cov} [ \hat{w}_{ij}(\theta),  \hat{w}_{mn}(\theta')] =  \sum_{\ell} \frac{(2 \ell +1)}{(4 \pi)^2 f_{\rm sky}   }   \bar{\mathcal{ L}}_\ell(\cos \theta) \bar{\mathcal{L}}_{\ell}(\cos\theta') \nonumber \\
\Big[ \Big( & C^{im}_{\ell}+ \frac{\delta^{im}_{\rm K}}{\bar{n}_i}\Big)   \Big( C^{jn}_{\ell}+\frac{\delta^{jn}_{\rm K}}{\bar{n}_j}\Big) + \Big(C^{in}_{\ell}+\frac{\delta^{in}_{\rm K}}{\bar{n}_i }\Big) \Big(C^{jm}_{\ell}+\frac{\delta^{jm}_{\rm K}}{\bar{n}_j}\Big) \Big], 
\end{align}
where $ \bar{\mathcal{ L}}_\ell $ represents the bin-averaged  Legendre polynomial \cite{Salazar-Albornoz:2016psd}, $f_{\rm sky} $ denotes the fraction of the sky coverage, and $ \delta_{\rm K} $ is the Kronecker delta.  The Poisson shot noise is assumed and  $\bar{n}_i $ is the angular number density in bin $i$.  We compute $C_\ell $ using the {\tt camb sources} code \citep{Lewis:2007kz}.


The theory template and the covariance described here are the key ingredients for the likelihood analysis discussed below.

\subsection{Parameter inference}

The correlation function measures the excess galaxy pair counts relative to the random pair counts.  We measure the spatial correlation function using the Landy-Szalay estimator \cite{LS1993}
\begin{equation}
\xi_{\rm p} (s_\parallel , s_\perp ) = \frac{ DD - 2DR +  RR }{RR },
\end{equation}
where $DD$, $DR$, and $RR$ denote the normalized pair counts of the data-data, data-random, and random-random pairs, and the results are binned in terms of the radial separation $s_\parallel$ and transverse separation  $s_\perp$.  The measurements are performed using the public code {\tt CUTE} \citep{Alonso_CUTE}.

In this work, we consider maximum  $ s_\parallel $  and  $ s_\perp $ up to $120 \MpcOh$ and $ 175 \MpcOh  $, respectively.   The measurements are further binned into $s$ and $\mu$.  The final $\xip$ is obtained by stacking the pairs together with some window function. Note that the maximum parameter  $ s_\parallel $ and $ s_\perp $  and the stacking window must be the same as those used in  the template computation.  


Under the Gaussian likelihood assumption, the likelihood $\mathcal{L} $  is 
\beq
\mathcal{L} \propto \exp \Big( - \frac{ \chi^2 }{ 2}  \Big)
\eeq
with $\chi^2 $ defined as 
\beq
\chi^2 = \sum_{ij} ( M_i - D_i) C^{-1}_{\phantom{...} ij} ( M_j - D_j) , 
\eeq
where $ M $ denotes the model vector, $D$ the data vector, and $C^{-1} $ the inverse of the covariance matrix.

The full model for the BAO fit is given by
\beq
\label{eq:full_nuisance_model}
M(s_\perp) = B \, T( \alpha s_\perp )  + \sum_i \frac{ A_i }{ s_{\perp}^i },     
\eeq
where $T$ signifies the theory template computed in the fiducial cosmology as described in Sec.~\ref{sec:theory}. The parameters in Eq.~\eqref{eq:full_nuisance_model} are elaborated below.

The dilation parameter $ \alpha $ enables us to shift the BAO position in the fiducial cosmology to match the one in the data cosmology. Because $\xip$ traces the underlying correlation at the transverse scale at the level of DES photo-$z$ uncertainty, we adopt  $s_\perp$ as the independent variable.  The measurement of $ \xip $ constrains the transverse BAO scale via
\beq
\label{eq:alpha}
\alpha \frac{r_{\rm s}  }{ D_{\rm M}( z_{\rm eff} ) } = \frac{r_{\rm s}^{\rm fid}  }{ D_{\rm M}^{\rm fid}( z_{\rm eff} ) },   
\eeq
where $r_{\rm s}$ and $  D_{\rm M} (z_{\rm eff} )$ denote the sound horizon at the drag epoch and the comoving angular diameter distance to the effective redshift of the sample, and ``fid'' signifies that the quantity is evaluated at the fiducial cosmology. At $z_{\rm eff} = 0.835$,  for the fiducial Planck cosmology, $ r_{\rm s} = 147.6 $ Mpc and $D_{\rm M} (z_{\rm eff}) = 2967.0 $ Mpc, while in MICE cosmology,  $ r_{\rm s} = 153.4 $ Mpc and $D_{\rm M} (z_{\rm eff}) = 2959.7 $ Mpc.

As in standard BAO analyses (e.g.~\cite{Xu_etal2012,Anderson_etal2014,Ross:2017emc}), extra parameters are introduced to accommodate the overall amplitude and shape of the correlation. The parameter $B$ allows for amplitude adjustment and the polynomial in  $ 1 / s_{\perp} $ is introduced to absorb the imperfectness in the modeling of the correlation function, shape changes due to difference in cosmology, and possible residual systematic correction. The default choice for $ A_i$ is $i=0$, 1, and 2.

To look for the best fit, we minimize $ \chi^2$  following the procedures in \cite{Chan:2018gtc}. We first fit the linear parameters $A_i$ analytically. Second, the residual $\chi^2$ is minimized with respect to $B$ numerically under the condition $B>0$. Finally, we search for the minimum of the resultant $\chi^2 $  with respect to $\alpha$. Note that the sequential search is adopted mainly for its speed and convenience, and it yields similar results as the MCMC fit, in which all the parameters are varied simultaneously.  We estimate the 1-$ \sigma $ error bar for $\alpha$ by applying the  $\Delta \chi^2 =1 $ criterion on the final residual $\chi^2$.

\begin{figure*}[!htb]
\centering
\includegraphics[width=\linewidth]{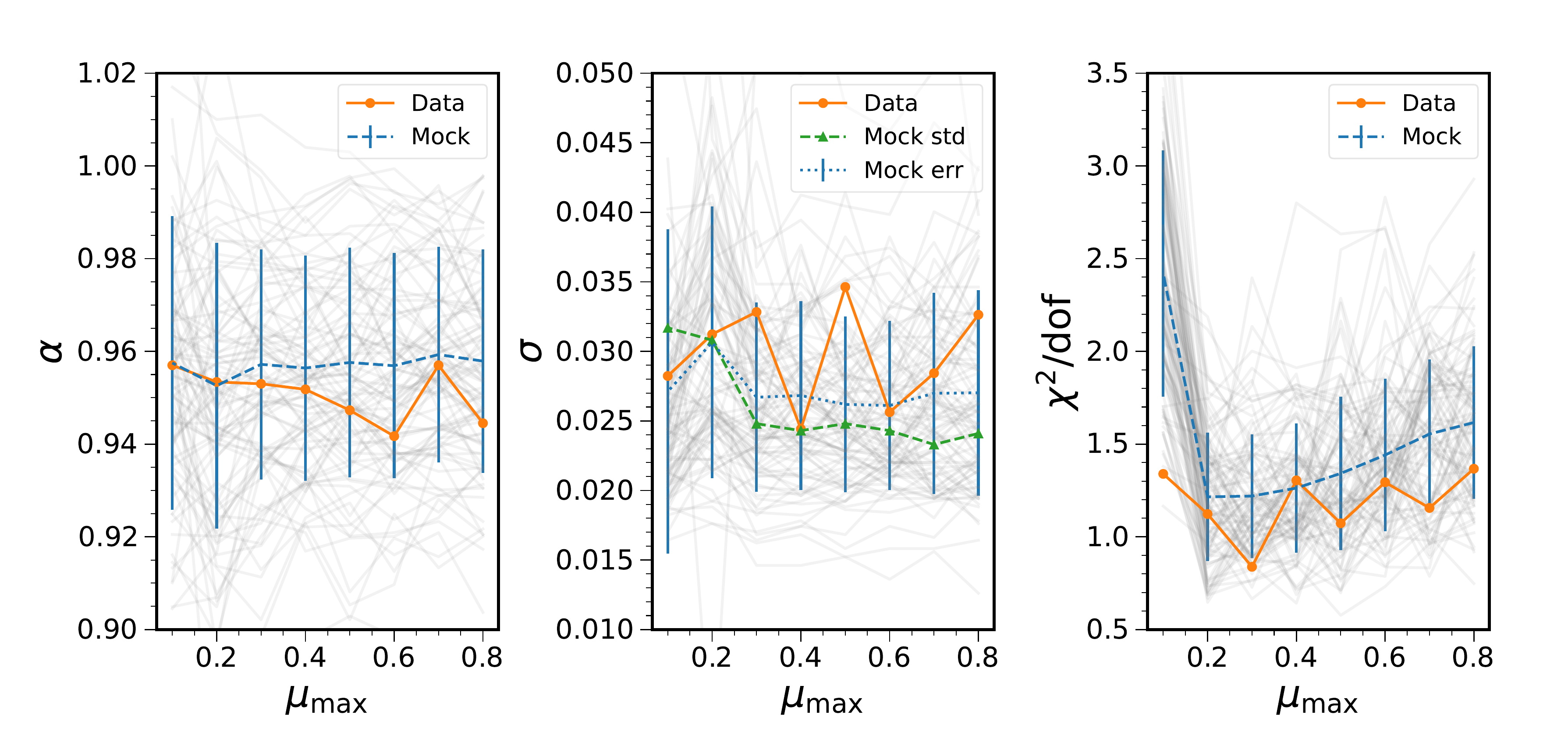}
\caption{  The BAO fit results obtained with the top-hat stacking window $ W_{\rm TH} ( \mu, \mu_{\rm max} )$ [Eq.~\eqref{eq:TH_stacking}] for different maximum cut-off $ \mu_{\rm max}$. The left, middle, and right panels show the best fit $\alpha$, the estimated error bar for the best fit, and the $\chi^2$ per degree of freedom, respectively.   The grey lines show the results from the individual mocks, and the blue curves with error bars indicate their mean and standard deviation.   The orange line corresponds to the actual data fit.  The middle panel also displays the standard deviation of the best fit (green).  Note that the best fit $\alpha $ is expected to cluster around 0.959  for the fiducial Planck template fit to the MICE mocks.     } 
  \label{fig:mock_data_mumax}
\end{figure*}

\begin{figure*}[!htb]
\centering
\includegraphics[width=\linewidth]{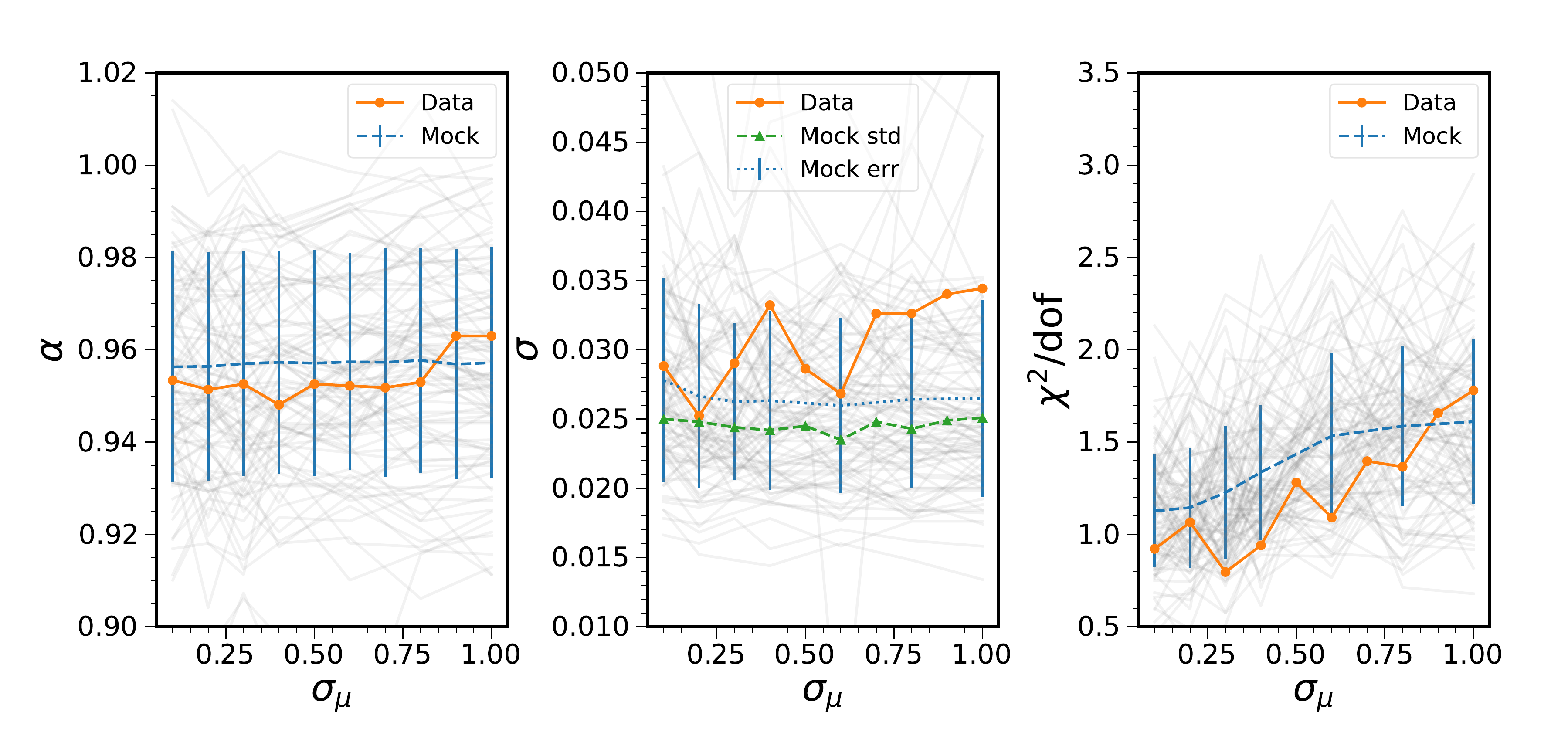}
\caption{  Similar to Fig.~\ref{fig:mock_data_mumax} but for the cut-off Gaussian stacking window $ W_{\rm G} ( \mu, \sigma_\mu )$ [Eq.~\eqref{eq:cutoffGaussian_stacking}], as a function of the dispersion of the window $\sigma_\mu$. The  $ W_{\rm G} $  window gives  more stable results than  $ W_{\rm TH} $ does.    }
  \label{fig:mock_data_Gweight}
\end{figure*}

\section{ Mock tests  }
\label{sec:mock_tests}

The  $ \xip $ method  has been extensively tested against a set of dedicated DES Y3 mocks, the ICE-COLA mocks \cite{DES:2021fie} in \cite{Chan_etal2021}.  In this work we shall present some further mock test results.

We first briefly describe the mock catalog, the ICE-COLA mocks, and refer readers to \cite{DES:2021fie} for more details.  The ICE-COLA mocks are generated from the COLA simulations which were run with the ICE-COLA code \citep{ice-cola} employing the COLA method \citep{Tassev_etal_2013}.  This method combines the second order Lagrangian perturbation theory with the particle-mesh simulation technique to make sure that the large-scale modes remains accurate when coarse simulation time steps are used.  The simulation consists of $2048^3$ particles in a cube of side length 1536 $\MpcOh$ so that its mass resolution coincides with that of the MICE Grand challenge $N$-body simulations \citep{Fosalba_etal2015,Crocce_etal2015}.  The mock galaxies are allocated to the halos using a hybrid Halo Occupation Distribution and Halo Abundance Matching recipe as in \cite{Avila:2017nyy}. The redshift distribution and the bias parameters follow the measurements of the actual data.  The same VIPERS dataset used to calibrate $ \phi $, also enables us to estimate a two-dimensional distribution $P(z_{\rm p}, z_{\rm s} )$, which is subsequently used to assign realistic photo-$z$ to the mock galaxies.  The simulation is replicated three times in each Cartesian direction (64 copies in total) to form a full sky lightcone mock up to $z \sim 1.4$.  
From each full-sky light-cone mock, four DES-footprint mocks are extracted.  
Due to limitation in computing power, about a hundred mocks are used, with the precise number depending on the test in question.

Before going over the test results, let us discuss the blinding policy in DES data analysis.  The aim of this practice is to prevent confirmation bias.  Before fixing the analysis pipeline, we are not allowed to look into the cosmologically interesting part of the data. A battery of pre-unblinding tests were devised  in DES Y3 to test the validity of the data and the soundness of the methodology without violating the blinding policy \cite{DES:2021esc}. Only after the tests are passed, the pipeline is fixed and the data are unblinded.


The initial phase of this project strictly followed the DES blinding protocol.  The pipeline for  $\xip$ is mainly guided by  the test results in \cite{Ross:2017emc,Abbott:2017wcz,Chan_etal2021}.  We had performed a battery of pre-unblinding tests similar to those in DES Y3. Many of them are similar to the robustness tests to be presented  in Sec.~\ref{sec:robustness_test}.   We initially adopted the stacking window $ W_{\rm TH} $ with $ \mu_{\rm max} = 0.8 $ following \cite{Ross:2017emc}.  However, after passing the pre-unblinding tests and unblinding, we realized that this choice was not ideal, and considered the Gaussian window as an alternative.  We find that $W_{\rm G}$ is more robust to analysis choices, and this will be evident below.  Although the adoption of the Gaussian window does not bias our results (will be clear later on), the Gaussian window results are not blinded  according to the blinding policy.  For completeness, we show the pre-unblinding test results in Appendix~\ref{sec:PreunblindingTest}.  Here we present the tests on the stacking window on the mocks.


\subsection{ Test of stacking windows  }

In this subsection, we test the results obtained with the top-hat window [Eq.~\eqref{eq:TH_stacking}] and the cut-off Gaussian window [Eq.~\eqref{eq:cutoffGaussian_stacking}].  In the following mock test results, the fiducial cosmology is assumed to be Planck even though the mocks are constructed in MICE cosmology.

In Fig.~\ref{fig:mock_data_mumax}, we first show the results for $ W_{\rm TH} ( \mu, \mu_{\rm max} )$ against $ \mu_{\rm max}$.  In the left panel, the best fit $\alpha$ from individual mocks, and their mean and standard deviation for different  $ \mu_{\rm max}$ are shown.   The best fit is approximately constant with similar spread for $ \mu_{\rm max} \gtrsim 0.3 $, below which the spread  starts to increase.   The increase in fluctuation can be attributed to the reduction in the data size as $ \mu_{\rm max} $ decreases.  In the middle panel, the error bars from the individual mock fit and their corresponding mean and standard deviation are plotted. As a comparison, the standard deviation of the best fit is also overplotted.  For the derived error bars, we find a similar trend that they increase as  $ \mu_{\rm max} $ decreases for  $\mu_{\rm max} \lesssim 0.3  $, consistent with that of the standard deviation.     While using  $\mu_{\rm max} \gtrsim 0.3  $ does not tighten the constraint on $\alpha$, it does cost a larger $\chi^2$.  On the right panel of Fig.~\ref{fig:mock_data_mumax}, we plot the $\chi^2$ per degree of freedom  ($\chi^2 / {\rm dof} $), which decreases as $ \mu_{\rm max} $ decreases up to $  \mu_{\rm max} = 0.2 $, below which it shoots up.    Photo-$z$ mixing in the radial direction caused the resultant covariance to be highly correlated \cite{Chan_etal2021}.  This poses difficulties for the  data analysis.  
Among them is that the $\chi^2 / {\rm dof} $ is substantially larger than 1 ($p$-value can be $8 \times 10^{-4} $) even though the fit appears to be good, i.e. the best fit is well within all the (correlated) 1-$\sigma$ error bars.  In \cite{Chan_etal2021}, using the orthogonal basis, it was shown that the BAO scale is well fitted by the model and the issue stems from the scale smaller than the BAO scale. Our results here further support that the issue of large  $\chi^2 / {\rm dof} $ originates from photo-$z$ mixing because reducing $\mu_{\rm max} $ decreases the impact of the photo-$z$ mixing.   However, for $ \mu_{\rm max} =0.1$,  the resultant $\chi^2 / {\rm dof} $ shoots up significantly. This coincides with the (less dramatic) increase in the fluctuation of the best fit. Reduction in the data size causes the likelihood to deviate from Gaussianity, violating the Gaussian likelihood approximation.     The mock test suggests that using $ \mu_{\rm max } \sim 0.3 - 0.4 $ is close to optimal for  $ W_{\rm TH}  $ because it does not weaken the constraint on $\alpha$ and still enjoy the benefit of low  $\chi^2 / {\rm dof} $.

We display the corresponding results for   $ W_{\rm G} ( \mu, \sigma_\mu )$ in  Fig.~\ref{fig:mock_data_Gweight}. This window gives more weight to the small $\mu$ pairs, and hence suppresses the high noise modes.
The best fit behaves stably with  $\sigma_\mu $. Small fluctuations are observed for the largest $\sigma_\mu$ shown, but these already include a significant fraction of the high  $\mu$ pairs. Shown in the middle panel are the estimated error bars. Similar to the best fit, the error bars are also stable w.r.t.~variation in   $\sigma_\mu$. Unlike the top-hat window case, in the smallest range shown  $\sigma_\mu \sim 0.1 $, there is only mild increase in the error bar size accompanying with tiny increment in the standard deviation of the best fit.   Similar to  $ W_{\rm TH} $ window case, we find that the $\chi^2 /{\rm dof} $ decreases as $ \sigma_\mu  $  decreases. However, it does not increase for  $ \sigma_\mu =0.1 $; instead it becomes saturated.

The mock test results demonstrate that adopting  $ \sigma_\mu \sim 0.2- 0.3 $  achieves a low value of  $\chi^2 /{\rm dof} $ without losing  parameter constraint. Moreover, $ W_{\rm G} $ is preferred to  $ W_{\rm TH} $ because the former gives more stable results.  In the following, we shall mainly discuss two cases: $ W_{\rm TH}$  with $\mu_{\rm max} = 0.8 $  for the ``historical reason'' and the optimal case  $ W_{\rm G}$  with $\sigma_{\mu} = 0.3 $. For convenience we will often simply abbreviate these two cases as  $W_{\rm TH} $ and $ W_{\rm G} $, respectively.  We note that the $ W_{\rm TH}$  with $\mu_{\rm max} = 0.8 $ results are  blinded in accordance with the blinding policy.  The precise reason for the choice of  $\sigma_{\mu} = 0.3 $  is that its error bar size for the actual data fit is close to mean error in the low  $\sigma_\mu$  regime.

\begin{figure*}[!htb]
\centering
\includegraphics[width=\linewidth]{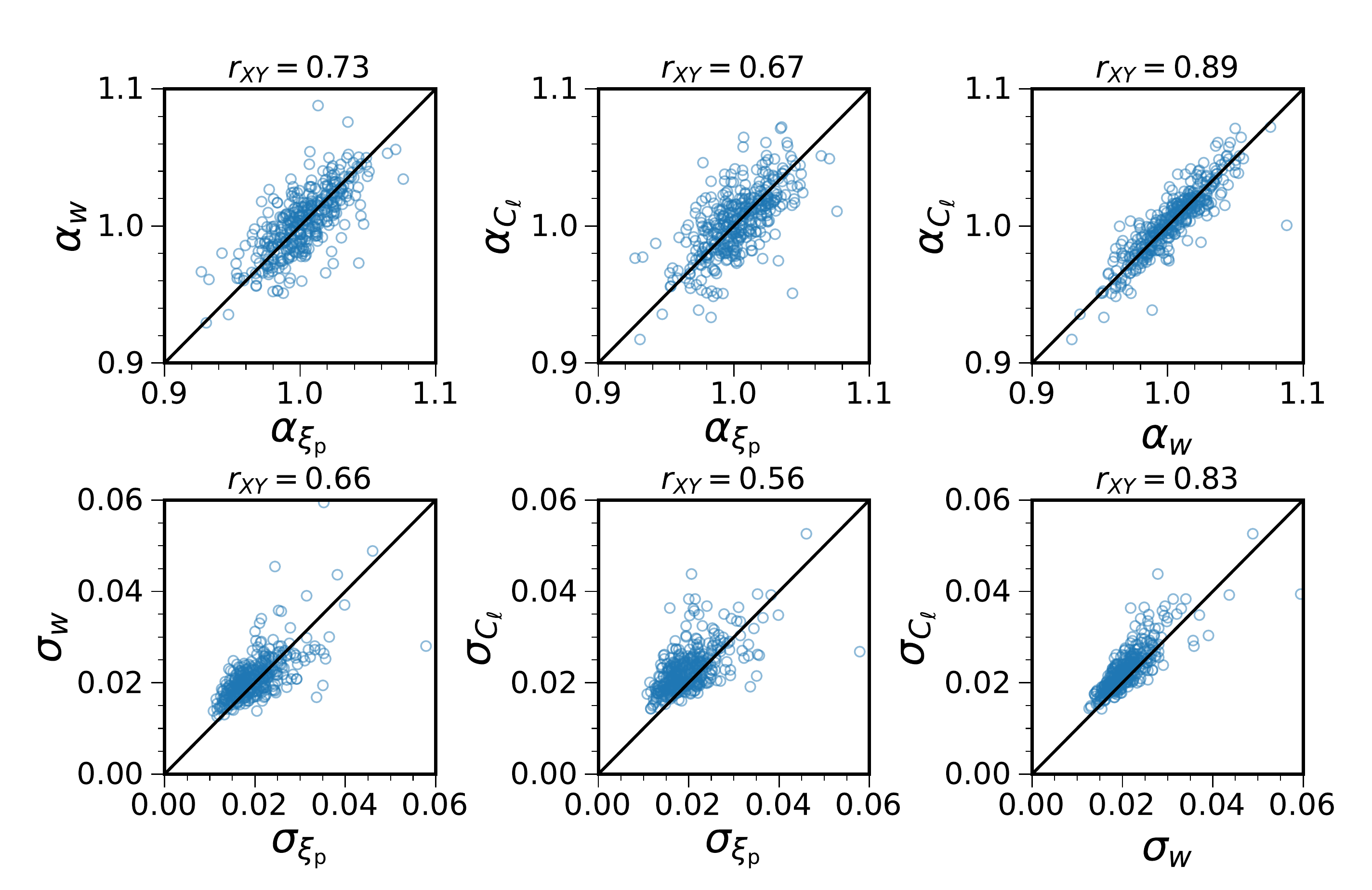}
\caption{  The scatter plot for the best fit $\alpha$ and the error bar $\sigma$ from the ICE-COLA mocks. The results obtained using $\xip$ ($ W_{\rm G} $),  the angular correlation function $w$, and the angular power spectrum $C_\ell$ are compared.  The best fit $\alpha$ from $\xip$ and $w$,  $\xip$ and $C_\ell$, and $w$ and $C_\ell$ are plotted in the upper panels, while the derived error bar $\sigma$'s ($\sigma_{\xip} $ versus $\sigma_w$,  $\sigma_{\xip} $ versus $\sigma_{C_\ell}$, and $\sigma_w$ versus $\sigma_{C_\ell }$) are shown in the lower panels. The Pearson correlation coefficients are also printed.  The best fit $\xip$ results are less correlated with $w$ or $C_\ell$ results relative to the those between $w$ and $C_\ell$.  The result are similar for $\xip$ with $ W_{\rm TH} $, but the correlation coefficients are slightly smaller.   }
  \label{fig:scatter_xip_w_Cl}
\end{figure*}

\subsection{ Correlation of the statistics }  

To better understand the $\xip$ statistics and to facilitate the comparison with the Y3 BAO results,  we compare the best fit results from $\xip$ against those derived from the angular correlation function $w$ and the angular power spectrum $C_\ell$ (see \cite{DES:2021esc} for the details on these measurements).  The best-fit $\alpha $ and the corresponding error bars from these statistics are compared in Fig.~\ref{fig:scatter_xip_w_Cl}.  The $\xip $ results are obtained using   $ W_{\rm G} $.    We find that there is larger scatter between $\xip$ and $w$ (or $C_\ell$) results relative to that between $w$ and $C_\ell$.

To quantify the correlation of the measurements, we use the Pearson correlation coefficient 
\beq
r_{ XY} = \frac{ \mathrm{cov}(X,Y)  }{ \sigma_{ X} \sigma_{\rm Y} },
\eeq
where $\mathrm{cov}(X,Y)$ is the covariance between $X$ and $Y$, and  $ \sigma_{ X}$ ($ \sigma_{\rm Y}$) is the standard deviation of $X$ ($Y$). The correlation coefficients are also shown in Fig.~\ref{fig:scatter_xip_w_Cl}. The correlation between $\alpha_{\xip}$ and  $\alpha_w$ ($\alpha_{C_\ell})$ is only 0.73 (0.67), and it is low compared to that between  $\alpha_w$ and $\alpha_{C_\ell}$, which reaches 0.89.  The correlation between the error estimates are 0.66, 0.56, and 0.83 for $ r_{\sigma_{\xip}  \sigma_w }$, $ r_{\sigma_{\xip}  \sigma_{C_\ell} }$,  and  $ r_{ \sigma_w  \sigma_{ C_\ell } } $, respectively.  These are generally smaller than those for the best fit values.   In contrast, for   $  W_{\rm TH}  $, we have $r_{\xip w  } = 0.70$, $r_{\xip C_\ell  } = 0.66$,  $r_{ \sigma_{ \xip} \sigma_w  } = 0.59$, and   $r_{ \sigma_{ \xip} \sigma_{C_\ell}  } = 0.50$, respectively.

It is easy to understand that $w$ and $C_\ell$ exhibit high level of correlation because both are the auto-correlation analysis of five tomographic bins, with the difference that one is in configuration space and the other in harmonic space.  On the other hand,  $\xip$ combines the information in five tomographic bins into a single data vector by including  all the correlation without explicit binning in redshift.   Moreover, they differ in the order of projection and correlation measurement. While the angular statistics first project the data to the angular space and then measure the correlation, $\xip$ goes the opposite way. We will argue that the difference in ordering  has important consequences on the stability of the estimator.   The correlation coefficients from   $W_{\rm TH}$  are lower than those of  $W_{\rm G}$, in agreement with the expectation that the high $\mu$ pairs give lower correlation with the transverse scale.   The fact that the $\xip$ measurements are less correlated with the tomographic angular analysis results implies that it can provide a relatively independent measurement and can offer an important crosscheck  because they could have different sensitivity to the potential systematics.



\section{Results}

\label{sec:results}

In this section, we first present measurements of the BAO with $\xip$ on the Y3 BAO sample, and then discuss the robustness tests performed to test the stability of the results. 

\subsection{BAO measurements}
\label{sec:BAOmeasurements}

\begin{table}[!tb]
\centering
\caption{ Constraints on the physical parameter $ D_{\rm M} /r_{\rm s} $ in Planck and MICE cosmologies. The all bin cases and the combo 2-4 bins cases are compared.     } 
\begin{tabular}{l  c|c|c }
  \hline 
  Case                 &   $\xip $:  $  W_{\rm G} $   &   $\xip $:  $ W_{\rm TH}  $         &  $w$     \\ \hline
  Planck (all bins)     &  $19.15 \pm 0.58  $                &    $19.00 \pm 0.67  $                      & $ 18.84 \pm 0.50 $    \\ 
  MICE (all bins)     &  $19.22 \pm 0.50  $                &    $19.15 \pm 0.42  $                      & $ 18.86 \pm 0.42 $    \\ \hline
  Planck (2, 3, 4 bins)     &  $19.84 \pm 0.53  $                &    $19.80 \pm 0.67  $                   & $ 19.86 \pm 0.55 $    \\ 
  MICE (2, 3, 4 bins)       &  $19.86 \pm 0.35  $                &    $20.12 \pm 0.39  $                   & $ 19.76 \pm 0.47 $    \\ \hline  
  \hline
  \label{tab:DMrs_constraint}
\end{tabular}
\end{table}

\begin{figure}[!htb]
\centering
\includegraphics[width=\linewidth]{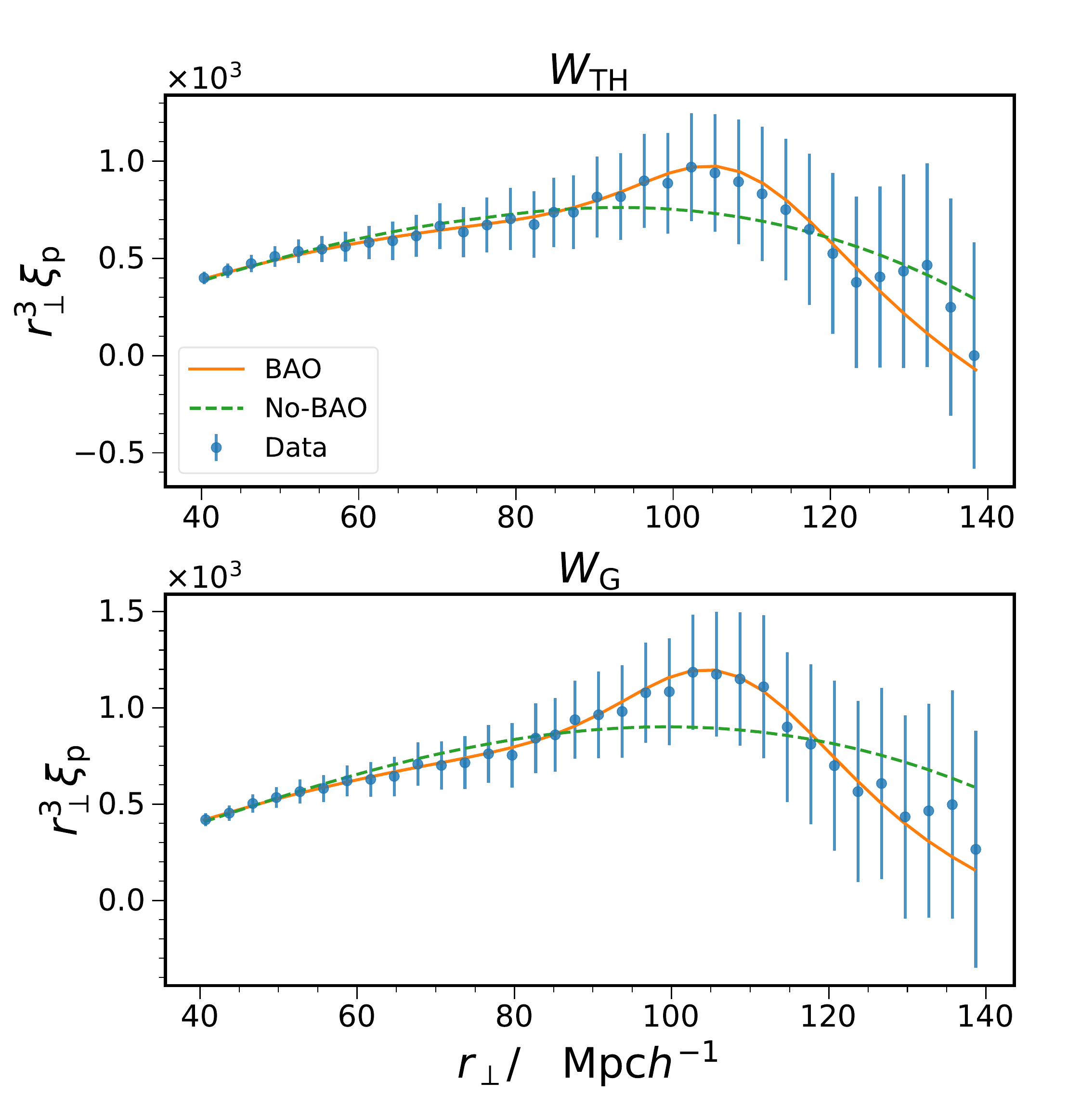}
\caption{  The measurement of $\xip$ (data points with error bars) and the best-fit model as a function of the transverse scale $ r_\perp = r \sqrt{ 1 - \mu^2 }$. The model  with the BAO feature (solid, orange) and without the BAO (dashed, green) are compared.  The results obtained with $W_{\rm TH} $ and $  W_{\rm G}  $ are displayed in the upper and lower panel respectively.     }
  \label{fig:xip_bestfit_WTH_WG}
\end{figure}

\begin{figure}[!htb]
\centering
\includegraphics[width=\linewidth]{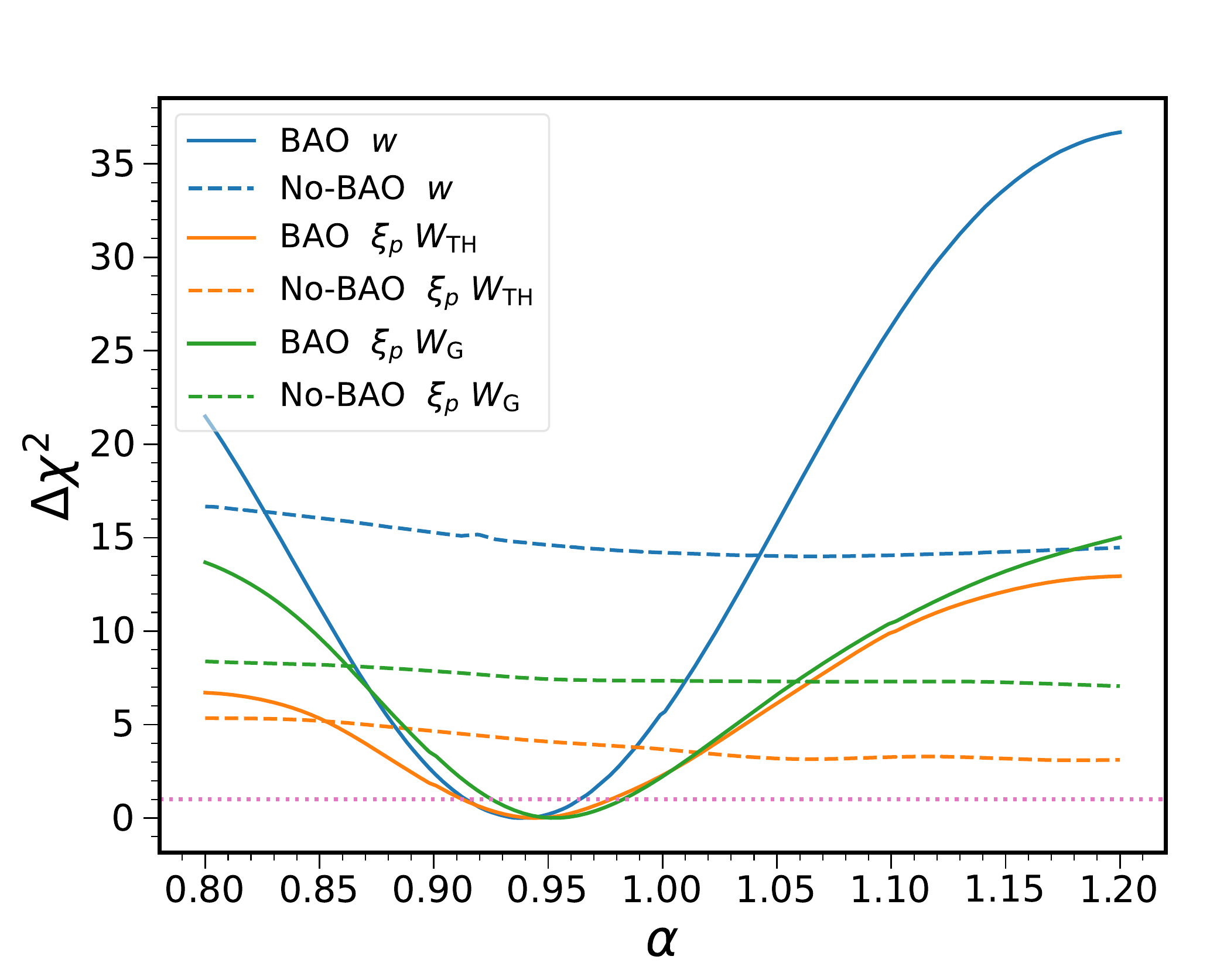}
\caption{ $ \Delta \chi^2 \equiv \chi^2 -\chi^2_{\rm min} $ as a function of $\alpha$. The BAO fit results from $ \xip$ (orange for $ W_{\rm TH} $ and green for $  W_{\rm G}  $) are compared with those from $w$ (blue).  The $ W_{\rm TH} $ results are  blinded, while  $  W_{\rm G}  $ ones are unblinded.  Both the results from the BAO template (solid) and the no-BAO template (dashed) are shown. In either case, $\chi^2_{\rm min}$ from the BAO template fit is subtracted.  The dotted red line indicates $\Delta \chi^2 =1 $, whose intersection with the $\Delta \chi^2$  curve gives the 1-$\sigma$  error bar.    Because the BAO signal in the sample is heterogeneous in redshift, the $\xip$ constraint from the combined sample is weaker than that from $w$ (c.f.~Fig.~\ref{fig:bestfit_Dchi2_bins2-4}).       }
  \label{fig:bestfit_Dchi2}
\end{figure}

Before presenting the BAO  measurements, let us turn to discuss the stacking windows.    As mentioned previously, initially we strictly abided by the blinding protocol, and the pipeline  was fixed based on the previous mock test results \cite{Ross:2017emc,Abbott:2017wcz,Chan_etal2021}.  In particular, the blinded pipeline uses the top-hat window with $ \mu_{\rm max} = 0.8 $. However, after unblinding,  we realized  that the stacking window can have a large impact on the results.   The effects of the stacking window are tested on the  mocks and the actual data, and the results are  shown in Figs.~\ref{fig:mock_data_mumax} and  \ref{fig:mock_data_Gweight}  for the top-hat and the Gaussian window, respectively.  Because the high-$\mu$ pairs are less correlated with the low-$\mu$ ones, the top-hat window results  are less stable w.r.t.~variation in $\mu_{\rm max} $.

This expectation is corroborated by the mock test results.  We also find a similar trend in the data. For the actual data fit with $W_{\rm TH} $, there are large fluctuations in the best fit value, and it becomes stable for $\mu_{\rm max} \lesssim 0.3 $. The resultant error bar from the data fit shows even larger fluctuations, and it only becomes relatively mild for $\mu_{\rm max} \lesssim 0.3 $. In this regime, however the mock test suggests that the spread increases as $\mu_{\rm max} $ decreases. The trend for the $\chi^2/{\rm dof} $ is similar to the mock result, i.e.~it decreases as  $\mu_{\rm max} $ decreases until   $\mu_{\rm max} \sim 0.1 $, where it starts to increase.  The Gaussian window offers higher stability than the top-hat. Except for the largest $\sigma_\mu$'s shown, the best fit $\alpha$ in the actual data fit is stable w.r.t.~$\sigma_\mu $. There are larger uncertainties for the error bar from the data fit. It shows substantial fluctuations for $\sigma_{\mu} < 0.5 $.  The $\chi^2/{\rm dof} $ shows a clear decreasing trend as $\sigma_\mu $ decreases, consistent with the trend found in the mock results. We adopt  $\sigma_\mu =0.3 $ because the estimated error bar is close to the average error bars in the range  $\sigma_{\mu} < 0.5 $.  It is worth emphasizing that although the Gaussian window is adopted after unblinding, the best fit is insensitive to the precise value of $\sigma_\mu $. This choice is also consistent with the recommendations we get from the mock test.

We now apply the fitting pipeline to the BAO sample and the results are shown in Fig.~\ref{fig:xip_bestfit_WTH_WG}, where we plot the measurement of $\xip$ and its best fit using both the template with and without BAO feature.  The results obtained with  $ W_{\rm TH} $ and  $ W_{\rm G} $ are visually similar.  The best fit $\alpha  $ is constrained to be $0.953 \pm 0.029 $ for  $  W_{\rm G} $, and   $\alpha =  0.945 \pm 0.033 $ for $ W_{\rm TH}  $.  In contrast, the angular correlation function yields  $ \alpha = 0.937 \pm 0.025 $.  Using Eq.~\eqref{eq:alpha}, we can translate them to the constraint on the physical parameter combination,  $ D_{\rm M} / r_{\rm s } $. Shown in Table~\ref{tab:DMrs_constraint} are  the results in Planck and MICE fiducial cosmologies.

 Table ~\ref{tab:single_bin_fit} displays  the $\chi^2 / \mathrm{dof}$ for the BAO template and the no-BAO template fit.  For $  W_{\rm G} $, the fit is very good with the BAO template yielding a  $p$-value of 0.84.  Although the $p$-value for the no-BAO template fit is also good (0.49), the BAO template results in a significantly smaller $\chi^2$.  For  $ W_{\rm TH}  $, the BAO template fit is decent with a $p$-value of 0.15 relative to the no-BAO template fit (0.08). For reference, the $p$-values for the $w$ fit are 0.31 (BAO) and 0.07 (no-BAO), respectively.   Fig.~\ref{fig:bestfit_Dchi2} shows the constraint on $\alpha$ by means of the $\chi^2$ values.   It displays $\Delta \chi^2$  as a function of $\alpha$, where $\Delta \chi^2 $ is defined as  $\Delta \chi^2 \equiv \chi^2 - \chi^2_{\rm min} $ with $\chi^2_{\rm min} $ being the minimum of $\chi^2 $. The 1-$\sigma$ error bar is given by the intersection of the $\Delta \chi^2 $ curve with the $ \Delta \chi^2 = 1$ horizontal line.  We also show the result obtained with the no-BAO template, for which we have subtracted the minimum of $\chi^2$, $\chi^2_{\rm min} $ from the BAO template fit. The difference between the minimum of the $\chi^2$ from the no-BAO template and BAO template can be used to claim the significance of the BAO detection.   As a comparison, the angular correlation function fit is overplotted.

We find that the error bar derived from $\xip$ is bigger than that from $w$ by a sizable amount.   Furthermore,  the $ \Delta \chi^2 $ between the BAO and the no-BAO model is much smaller than the $w$ result indicating that the significance of the BAO detection is lower.    By contrast, the error bars from the angular statistics are quite similar, with  $ \alpha = 0.937 \pm 0.025$ from $w$ and  $0.942 \pm 0.026$ from $C_\ell$.  While the mock tests suggest that $ \xip $ is likely to  yield a  slightly more competitive constraint than $w$ on average, Fig.~\ref{fig:scatter_xip_w_Cl} reveals that there is a significant fraction of mocks with  $ \sigma_{ \xi_{\rm p} } >  \sigma_w  $ and the correlation between $\xip$ and the angular statistics are not strong. Nonetheless, the somewhat weak constraint from $ \xip $ is worth further exploration.


\begin{figure}[!htb]
\centering
\includegraphics[width=\linewidth]{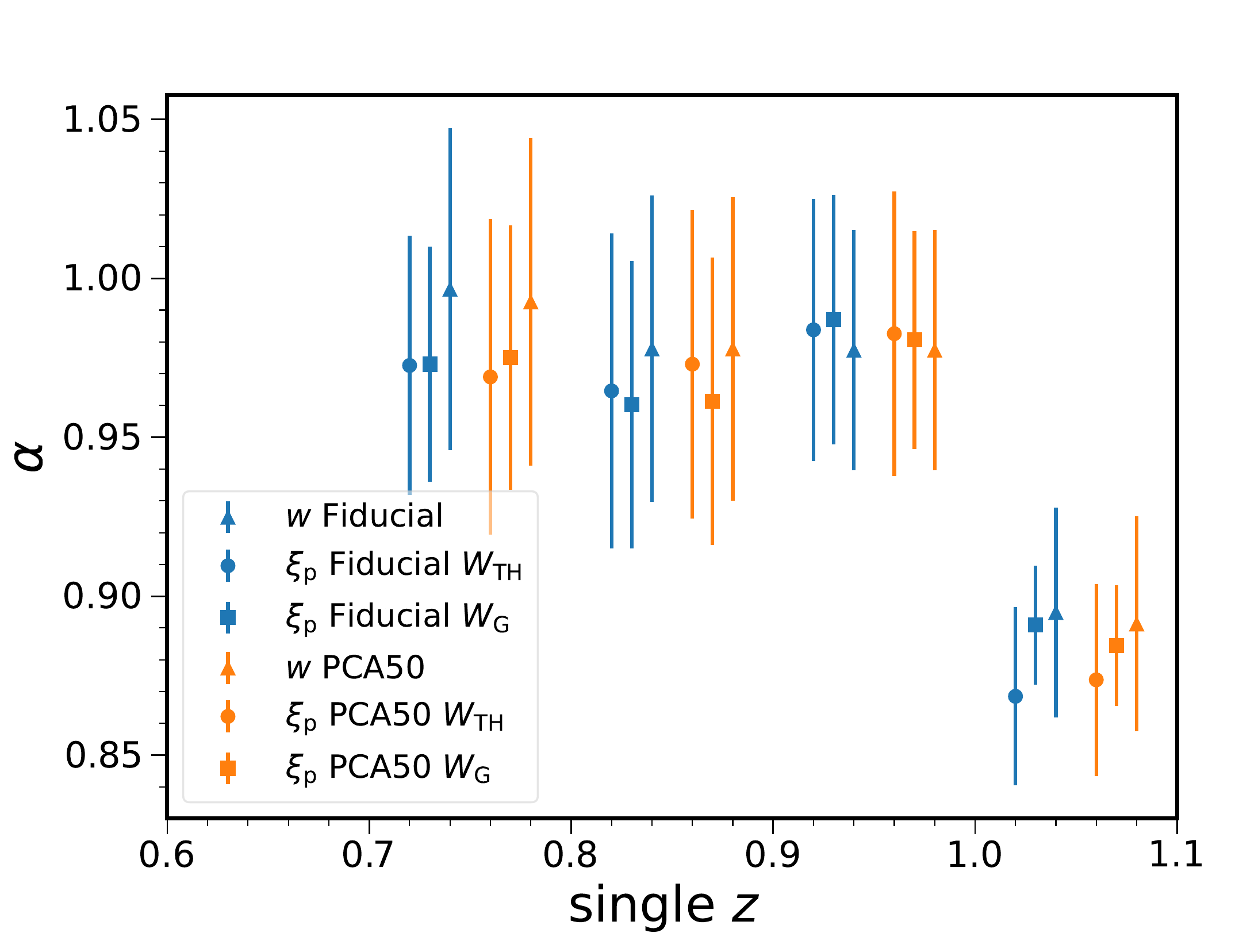}
\caption{  The BAO fit constraint on $\alpha$  obtained with data in a single redshift bin of width $\Delta z = 0.1$. The results for $\xip $ (circles for $ W_{\rm TH}  $ and squares for $  W_{\rm G}  $) are contrasted with those for  $w$ (triangles).  Note that there is no BAO detection for the first bin. Both the results obtained with the fiducial weight (blue)  and the alternative PCA50 weight (orange) are compared. The results from $\xip $ and $w$ are consistent with each other. }
  \label{fig:single_bin_fit}
\end{figure}

\begin{figure}[!htb]
\centering
\includegraphics[width=\linewidth]{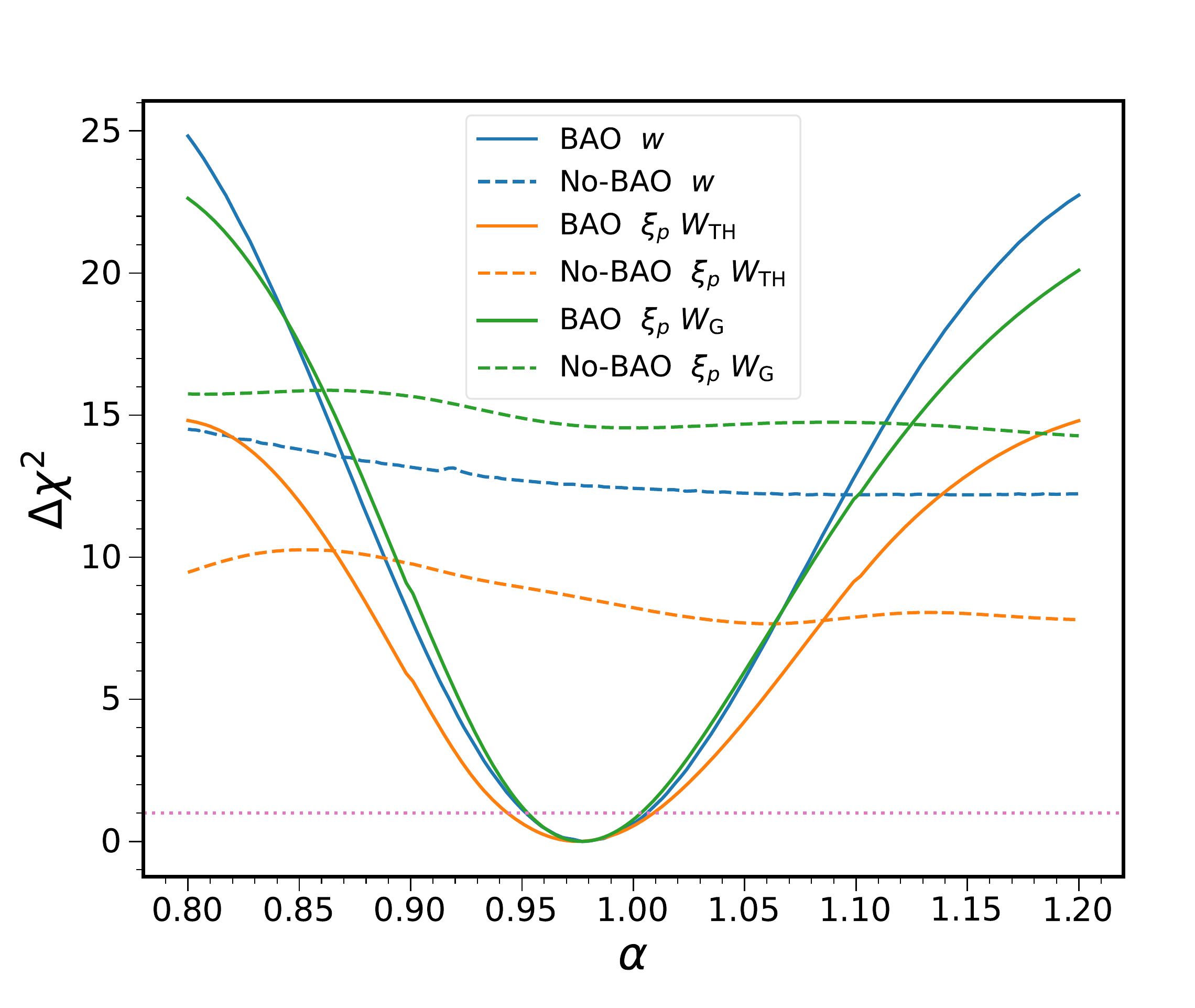}
\caption{ Similar to Fig.~\ref{fig:bestfit_Dchi2}, but for the combo 2-4 bins, whose BAO signals are homogeneous across redshift. For this sample, $ \xip $ with the Gaussian window $ W_{\rm G} $ yields the best constraint.      }
  \label{fig:bestfit_Dchi2_bins2-4}
\end{figure}

To shed light on this intriguing result, we look at the BAO fit on individual tomographic bin data.  The individual bin results are shown in Table \ref{tab:single_bin_fit} and visually in Fig.~\ref{fig:single_bin_fit}.  While there is no detection of BAO for the first bin, the signals are measured in all other redshift bins.  BAO is considered to be non-detectable if the best fit 1-$\sigma$ interval for $\alpha$ does not fall entirely within the interval [0.8,1.2].  The last bin shows unusually large deviation from the rest of the bins,  and it is responsible for the overall deviation from the Planck cosmology in DES Y3.

For single tomographic bin fit, $\xip$ with $  W_{\rm G}  $ overall yields  the smallest error bars and the size of the error bars from  $W_{\rm TH} $ are  similar to those from $w$.   This raises the question why for the full dataset, $\xip$ gives a weaker constraint.  Notice that the BAO information in the first and fifth bin are distinct from the rest, and this implies that the combined data are heterogeneous in terms of the BAO signals. \change{ The BAO signals should be constant according to standard model, however, due to random fluctuations, the measured signals could be heterogeneous across redshift.  A more pernicious cause for the heterogeneity in signals is some untreated systematics.  Here we emphasize that among the tests performed, there are no evidences suggesting that the heterogeneity in signals is caused by systematics.   }

Because the bin 2, 3, and 4 share similar BAO signals, it is illuminating to consider the BAO results for these bins combined, which are also shown in Table \ref{tab:single_bin_fit}.  The effective redshift of this homogeneous BAO-signal sample is 0.845. The $\chi^2 $ fit results for this sample are plotted in Fig.~\ref{fig:bestfit_Dchi2_bins2-4}.   We find that the best fit values are very consistent with each other.  For $ W_{\rm G} $ and $w  $, the goodness of the fit is broadly similar to the all bin case, but the   $W_{\rm TH} $ fit is significantly worse ($p$-value of 0.01). In this case, $\xip$ with $  W_{\rm G} $ yields an error bar of 0.026,  smaller than the all bin results by 0.03. For  $ W_{\rm TH} $, the error bar size is 0.33, the same as the all bin case. On the other hand, $w$ yields an error bar of 0.027, larger than the all bin case (0.025). Thus this test reveals that the $\xip$ statistic constraint is deteriorated if the signals in the sample are heterogeneous.  The degradation is less severe for the Gaussian window.  The angular statistics, however, still yield a tight bound in the presence of the heterogeneous signals.

In Appendix~\ref{appendix:errorbars}, we look into this issue further using the mock catalog. We select a sub-sample with heterogeneous BAO signals from the mock, by applying the criterion that the standard deviation of the best fit $\alpha$ among the five individual bins is larger than certain threshold.  Motivated by the actual data results, the threshold is set to be 0.07. This threshold  also balances with the number of mocks available.  We compute the probability that $\xip$ yields a larger error bar than $w$ for the whole sample and the heterogeneous sample.   Indeed,  we find that the probability of getting $\xip$ with error bar larger than that of $w$ becomes more appreciable for the heterogeneous mocks. This further supports the idea that $\xip$ yielding a larger error bar on the full sample is driven by the heterogeneity of the BAO signals.

In Table \ref{tab:single_bin_fit}, we also show the $\chi^2$/dof obtained with the BAO template and the no-BAO one.  For the no-BAO template, we choose the lowest $\chi^2 $ in the range [0.8,1.2].   For the full sample, the $ \Delta \chi^2 $ between the no-BAO template and the BAO one   for $\xip$ with either $ W_{\rm TH}  $ ($\Delta \chi^2 = 3.1$) or $ W_{\rm G} $ (7.1) are  substantially lower than that for $w$ (14.0). Thus for the full sample, the detection significance of the BAO from $\xip$ is significantly lower. However, for the fit on individual bins, we find that $  W_{\rm G}  $ yields a significantly higher  $\Delta \chi^2 $ and  $ W_{\rm TH} $ yields a similar   $\Delta \chi^2 $ relative to the corresponding $w$ result.  For the combo 2-4 bins,  $  W_{\rm G}  $ results in the highest $ \Delta \chi^2 $ (14.3), $w$ the second (12.2), and $ W_{\rm TH} $ the lowest (7.6).  Overall, we find that if the BAO signal in the sample is homogeneous,  $ W_{\rm G} $ gives the highest detection significance.

Their different response to the heterogeneous signals in the data is due to the ways that signals in tomographic bins are combined.  $\xip$ is measured by averaging the correlation signal in the whole sample and  the signals from the bins are combined to form a single data vector, and so the total BAO signal is smeared out if the signals in different sub-samples are not similar.  On the other hand, for $w$, the signals in the tomographic bins are combined at the likelihood level. It is easy to see this if the covariance between different bins can be neglected. The constraint always tightens when the likelihoods are combined.   In the extreme, it is well-known that when the likelihoods from inconsistent datasets are combined, the resultant constraint is artificially stringent. In this case, if some hyperparameters are introduced to model the systematics, the constraint will be loosened \cite{Hobson:2002zf,LuisBernal:2018drn}.   Thus the fact that $ \xip$ yields a weaker constraint than $w$ for heterogeneous signals does not necessarily mean that $ \xip$ is inferior compared to $w$.  On the bright side, $ \xip$ is capable of detecting the potential inconsistency in the dataset and reflects this in a poor constraint.  Then it boils down to whether the heterogeneous signals are genuine or not. The consistency of the data can be quantified by various tension metrics, e.g.~\cite{Charnock:2017vcd,Raveri:2018wln,DES:2020hen}, which can serve as diagnostic of potential systematics but not solution.   We note that there is no evidence suggesting that the heterogeneous signals in the full sample are caused by systematics.

\begin{table*}[!htb]
\centering
\caption{ Comparison of $\xip$ ($  W_{\rm G}  $  and   $ W_{\rm TH}   $) and $w$ BAO fit results on individual tomographic bins and other combinations.  Each tomographic bin is of width  $\Delta z_{\rm p} = 0.1$. The best-fit result and its corresponding  $\chi^2 $/dof (in parentheses) are shown. Besides, the  $\chi^2 $/dof for the no-BAO template fit  is shown in square brackets. Although for the full sample,  $\xip$ with $  W_{\rm G} $ yields a weaker constraint than $w$, it gives a tighter bound if the BAO signal in the sample is homogeneous as in the case of individual bins and the combo 2-4 bins.    } 
\begin{tabular}{l  c|c|c }
  \hline \hline
  Case             &   \multicolumn{3}{c}{ Fit Results  \,  ($\chi^2 $/dof ); \, [ $\chi^2 $/dof for the the no-BAO fit ]   }        \\ \hline
                   &   $\xip $:  $  W_{\rm G} $   &   $\xip $:  $ W_{\rm TH}   $         &  $w$     \\ \hline
  Bin 1 only       &  No detection      & No detection    & No detection                    \\
  Bin 2 only       &  $0.973 \pm 0.037 \,  ( 22.4/29 ) ; \,  [ 28.8/29]  $  &   $0.973 \pm 0.041 \,  ( 30.7/29 )  ; \,  [ 34.2/29]  $     & $ 0.997 \pm 0.051  \, (13.7/17) ; \,  [ 16.8/17]   $   \\
  Bin 3 only       &  $0.960 \pm 0.045 \, ( 51.6 /29 ) ; \,  [ 58.1/29]   $  &   $0.965 \pm 0.050 \, ( 42.6 /29 ) ; \,  [ 47.2/29]  $     & $ 0.978 \pm 0.048  \, (16.7/17) ; \,  [ 21.3/17]    $  \\
  Bin 4 only       &  $0.987 \pm 0.039 \, (  28.8/29 ) ; \,  [ 37.8/29]   $  &   $0.984 \pm 0.041 \, (  39.9/29 ) ; \,  [ 48.2/29] $     & $ 0.977 \pm 0.038 \,  (23.1/17 ) ; \,  [ 29.3/17]   $   \\
  Bin 5 only       &  $0.891 \pm 0.019 \,  ( 21.7/29 ) ; \,  [ 36.8/29]   $  &  $0.869 \pm 0.028 \,  ( 31.1/29 ) ; \,  [ 44.5/29]  $     & $ 0.895 \pm 0.033 \, (10.1/17 ) ; \,  [ 20.3/17]    $   \\  \hline
  Bins 2, 3, 4     &  $0.977 \pm 0.026 \, (22.9/29 ) ; \,  [ 37.2/29]     $  & $0.975 \pm 0.033 \,   (49.3/29 )  ;  \,  [56.9/29]  $     & $ 0.978 \pm 0.027 \, ( 53.9/53 ) ; \,  [ 66.1/53]   $   \\  \hline
  All bins         &  $0.953 \pm 0.029 \, ( 21.5/29 ); \,  [ 28.6/29]    $  &  $0.945 \pm 0.033 \, (  36.9/29 )  ; \,  [ 40.0/29]  $    & $ 0.937 \pm 0.025 \, (95.2 /89)  ; \,  [ 109.2/89]   $   \\
  \hline
  \hline
  \label{tab:single_bin_fit}
\end{tabular}
\end{table*}

\subsection{Robustness tests}
\label{sec:robustness_test}


\begin{table*}[!htb]
\centering
\caption{ Numerous robustness tests are conducted to check the stability and soundness of the results. $\xip$  with $  W_{\rm G} $ and $ W_{\rm TH}   $  are compared with $w$ results. The tests that are part of the pre-unblinding tests are indicated with a star.     } 
\begin{tabular}{l  c|c|c }
  \hline 
   Case                      &   $\xip $:  $  W_{\rm G} $   &   $\xip $:  $ W_{\rm TH}  $         &  $w$   \\
  \hline
  Default               &     $0.953 \pm 0.029$   $(21.5/29)$   &     $0.945 \pm 0.033$  $(33.4/29)$            &   $0.937 \pm 0.025$  $(95.2/89)$   \\
  \hline
    No  sys.~corr.       &   $0.942 \pm 0.029$   $(39.7/29)$     &    *    $0.938 \pm 0.033$   $(46.4/29)$       &   $0.935 \pm 0.026$  $(94.6/89)$   \\
  ${\rm sys-PCA50}$     &   $0.945 \pm 0.029$   $(22.8/29)$     &        $0.943 \pm 0.028$   $(36.0/29)$       &   $0.937 \pm 0.025$ $(94.9/89)$     \\
    $n(z)$ \ZMC           &    $0.948 \pm 0.029$  $(21.6/29)$     &  *      $0.943 \pm 0.034$   $(33.6/29)$       &   $0.935 \pm 0.025$  $(95.6/89)$  \\
     MICE template        &   $0.989 \pm 0.038$   $(53.5/29)$     &  *      $0.988 \pm 0.032$   $(78.5 /29)$      & $ 0.980 \pm 0.026  $  $(95.1/89)$   \\
     MICE cov.            &   $0.956 \pm 0.021$   $(23.7/29)$    &   *     $0.955 \pm 0.025$   $(41.0/29 )$      &  $0.936 \pm 0.021 $  $(125.8/89)$     \\
   MICE cosmology          &  $0.996 \pm 0.026$    $(59.3/29)$    &        $  0.995 \pm 0.021 $ $(90.7/29)$       & $0.977 \pm 0.022 $ $(125.8/89 )$  \\
   Unmodified cov.      &    $0.956 \pm 0.030$  $(21.3/29)$     &        $0.953 \pm 0.035$   $(32.7/29)$       &     ---              \\   
   $ [70, 130] \MpcOh $ &   $0.955 \pm 0.030$  $(11.7/16) $     & $0.965 \pm 0.031$   $(17.1/16 )$             &    ---                 \\
$ \Delta r = 5 \MpcOh$  &  $0.953 \pm 0.030$  $(19.1/15)$       &  $0.953 \pm 0.036$   $(16.2/15)$             &    ---                 \\
$ \Delta r = 2 \MpcOh$  &   $0.949 \pm 0.028$   $(38.1/44)$     &  $0.941 \pm 0.031$   $(44.5/45)$             &    ---                 \\
    No bin 1             &   $0.976 \pm 0.024$   $(29.5/29)$     & * $0.960 \pm 0.030$   $(38.7/29)$                &  $0.948 \pm 0.026$  $(67.8/71)$    \\ 
     No bin 2             &   $0.928 \pm 0.034$   $(19.0/29)$     & *  $0.931 \pm 0.034$   $(32.4/29)$               &   $0.929 \pm 0.026$  $(80.7/71)$   \\ 
     No bin 3             &    $0.938 \pm 0.034$   $(27.0/29)$    & * $0.941 \pm 0.038$   $(38.7/29)$                &   $0.935 \pm 0.028$  $(78.4/71)$   \\ 
     No bin 4             &   $0.928 \pm 0.033$   $(24.7/29)$     & * $0.943 \pm 0.034$   $(38.8/29)$                &  $0.925 \pm 0.028$   $(70.0/71)$   \\ 
     No bin 5             &   $0.950 \pm 0.030$   $(21.5/29)$     & * $0.959 \pm 0.029$   $(40.6/29)$               &  $0.967 \pm 0.026$  $(82.3/71)$    \\
  \hline
  \label{tab:robustness}
\end{tabular}
\end{table*}

In this subsection, we conduct various robustness tests to check the validity and soundness of the results.  The results are presented in Table ~\ref{tab:robustness}. Analogous test results from angular correlation function in DES Y3 are also reproduced here to facilitate comparison. Many of the robustness tests are similar to the pre-unblinding tests and they are indicated with a star in Table ~\ref{tab:robustness}.


{\bf  $\bullet$ Impact of systematics correction} The observational systematic effects are corrected by the systematic weights to avoid contamination of the cosmological results.  Recall that the fiducial systematic weights are assigned iteratively until the galaxy density does not show appreciable dependence on the survey properties. Testing of the impact and the effectiveness of the systematic weights on the mocks have been presented in \cite{DES:2021jns}. Here we test the impact of the systematic weights on the data measurement.

When there is no systematic weights applied at all, $\alpha $ is measured to be $0.942 \pm 0.029 $ for $  W_{\rm G} $ and  $0.938 \pm 0.033 $ for $  W_{\rm TH}  $.  While there is no change in error bar relative to the default value,  a  shift in the best fit value by -1.1\% and -0.7\%, respectively are observed.  For $w$ there is  only a shift in the best fit by 0.2\% although it also accompanies with a change in error bar size by 4\%.

There is an alternative means to derive the de-contamination weights using the principle components of the survey properties as the input systematic maps. The end product is another set of weights, referred to as PCA50  (see \cite{DES:2021bat}). We have compared the single bin fit results for these two types of weights in Fig.~\ref{fig:single_bin_fit}. The change in the best fit for bins 2, 3, 4, and 5 are -0.4\%, 0.9\%, -0.1\%, and 0.6\% for  $ W_{\rm TH}  $ and 0.2\%, 0.1\%, -0.6\%, and  -0.7\%  for $ W_{\rm G} $ respectively. The corresponding change in the error bars for these bins are 21.6\%, -2.0\%, 8.7\%, and 7.9\% for  $ W_{\rm TH}  $ and 12\%, 0.0\%, -12.8\%, and 1.1\% for  $  W_{\rm G} $. We find that the maximum percentage changes in both the best fit and the error bar for   $ W_{\rm G}  $ are less than those for $ W_{\rm TH} $.   In contrast, $w$ again shows remarkable insensitivity to systematics treatment with the maximum change in best fit less than 0.4\%  and maximum change in error bar less than 2.4\%.

We can understand why  $\xip$ is more sensitive to the systematic correction weights from its effects on the density field.  The systematic correction weights modify the density field in the radial direction  and the angular position.  For $w$, only the angular density is affected as the radial direction has been projected out, while for $\xip$, the weights affect both the radial density and the angular one.

{\bf  $\bullet$ Impact of the true redshift distribution} 
To compute the theory prediction, we need the conditional weighted true redshift distribution $\phi$ given by Eq.~\eqref{eq:conditional_true_z}.  We consider  $\phi$  estimated using the \ZMC output in the DNF algorithm. This serves as a cross check on the fiducial conditional true $z$ distribution derived from the VIPERS sample.  The percentage change in the best fit $\alpha$ is -0.5\%, -0.2\%, and -0.2\% for  $ W_{\rm G} $,  $ W_{\rm TH} $, and $w$ respectively. For the error bar,  $ W_{\rm TH} $ yields change by a few percent and others are unchanged. Thus the true redshift distribution is not a major concern.

{\bf  $\bullet$ Alternative fiducial cosmology } In the fiducial analysis, the Planck cosmology is adopted. Here we consider using the alternative MICE cosmology. This affects the cosmology used to compute the template and the covariance on the theory side and to perform the pair counts for the data measurement. Because of the difference in shape and amplitude of the correlation function, the bias parameters are different in these cosmologies. We show the results obtained with the MICE template, the MICE covariance, and MICE cosmology, by which we mean both the MICE template and  the MICE covariance are used.  We note that there is an additional layer of cosmology dependence in $ \xip$ relative to $w$ as in the data measurement a fiducial cosmology is necessary to convert angles and redshifts to distances. This cosmology is taken to be the same as that of the template.

The best fit values for  $\xip$ with  $ W_{\rm G} $ and  $ W_{\rm TH}  $ are similar. When the MICE template is used, the error bar for   $ W_{\rm TH}  $ and $w$ show little changes, but that for  $ W_{\rm G} $ is significantly inflated. We further note that $ \chi^2 $ increases substantially for $\xip$.   For the MICE covariance case, for $w$  the best fit is little affected (-0.1\%), while  $\xip$ with  $ W_{\rm G} $ and  $ W_{\rm TH} $   show  changes by 0.3\%  and 1.1\% respectively. In MICE covariance, the error bars are reduced in all cases, with $w$ and $ W_{\rm G} $  yield the same error bar.  Finally for  MICE fiducial cosmology,  the constraints on the physical parameter  are  shown in Table ~\ref{tab:DMrs_constraint}.  In MICE cosmology, the best fit increases slightly, but the error bar are reduced by a significant amount.  The percentage changes in the best fit are 0.4\% ($ W_{\rm G} $), 0.8\% ($ W_{\rm TH}  $), and 0.1\%  ($w$), and -14\%, -37\% and -16\% in the error bar, respectively.

Overall, $ w$ is the most robust to changes in fiducial cosmology. $\xip$ with  $ W_{\rm G} $  is generally more stable than  $ W_{\rm TH}  $, but it sometimes still shows large fluctuations such as in the MICE template case.

{\bf  $\bullet$ Variation in the fitting conditions }    In \cite{Chan_etal2021}, a couple of issues related to the highly correlated covariance were discussed. Some of the fit results are manifestly bad because the best fit completely fall above (or below) all the data points. The problem was alleviated by suppressing the largest eigenvalues in the correlation matrix. We have adopted the prescription as the fiducial setup. We show the results for the original unmodified covariance, the percentage change in the best fit and error bar are 0.1\% and 3.4\% (0.8\% and 6.1\%) for $ W_{\rm G} $ ($ W_{\rm TH}  $).

The default fit range is [40,140] $\MpcOh$.  We show the results for a narrower range, [70,130] $ \MpcOh$. For a narrower range, while  $ W_{\rm G} $  seems to remain at the same best fit position with the error bar slightly loosened,  $ W_{\rm TH} $ is quite different from the fiducial one. 

We also test the bin width dependence. The default value is $\Delta r  =3 \, \MpcOh$. For $ \Delta r = 5 \, \MpcOh $,  the percentage change in the best fit and error bar are  0.0\% and 3.4\% (0.8\% and 9.0\%) for  $  W_{\rm G} $ ($ W_{\rm TH}  $).  For $ \Delta r  = 2 \, \MpcOh$,  the percentage change in the best fit and error bar are  -0.4\% and -3.4\% (-0.4\% and -6.1\%) for  $ W_{\rm G} $ ($W_{\rm TH} $).  

These test results again show that $\xip$ with  $ W_{\rm G} $ is more robust than $ W_{\rm TH} $.

{\bf  $\bullet$ Missing bin test }  In this test, data in one of the tomographic bins is removed.  It is easy to understand this missing bin test by referring to the single bin results in Table \ref{tab:single_bin_fit} (or Fig.~\ref{fig:single_bin_fit}).   The first bin is unusual because there is a trough at the anticipated BAO peak position, but there is large bump at the scale much larger than the expected, driving  $ \alpha $ to a very small value, which is so small that it does not meet the detection criterion.  Removing this bin, all the best fit $\alpha$ increases. For $ \xip $, as the first bin contributes ``signal'' very different from the rest, removing it actually tightens the bound.   However, $w$ always gives a tighter bound when the data size increases  because it effectively combines the likelihoods from different bins.  The behavior of removing the second, the third, or the fourth bin are similar as they contribute similar signals. The best fit from  $ W_{\rm G} $ is similar to the $w$ results, and the results from  $ W_{\rm TH} $ is slightly more different from the others. We find that the error bars generally increase in all cases.  
Removing the fifth bin impacts the three statistics more disparately.  The best fit $ \alpha $ for   $ W_{\rm TH}  $ and $w$  both increase, but the error bar size for   $ W_{\rm TH} $ decreases and that of $w$ increases.  On the contrary, the best fit for  $  W_{\rm G} $ actually decreases. Judging from the best fit and the $\chi^2 $, removing the fifth bin does not affect the likelihood much for $  W_{\rm G}  $.  These highlight that combining heterogeneous signals at the data vector level can be tricky and non-intuitive, while the combining likelihood is relatively straightforward.


In summary, from these tests we find that $w$ is the least sensitive to changes in the fitting conditions, $ W_{\rm G} $ the second, and  $ W_{\rm TH} $ the most sensitive.  A  main difference between $\xip$ and  the angular statistics such as $w$  is the order of the projection and the correlation measurement.  For angular statistics, we first project and then do the correlation measurement, while  $\xip$ goes the other way.  Because photo-$z$ errors affect only the radial direction, by projecting the field to the angular space first, the photo-$z$ errors can be nulled to a large extent, and the subsequent angular correlation measurement is little contaminated by the noise owing to the photo-$z$ errors.  $\xip$ aims to keep some radial information, and so the correlation measurement is first performed and it is projected to the transverse direction afterwards.  However, this approach comes with the price that the photo-$z$ noise can sneak in to contaminate the correlation function. The noise causes $\xip$ to be less stable.  The Gaussian window with $  W_{\rm G} $ reduces the weight of the pairs with large $\mu$, and this can limit the impact of the photo-$z$ contamination making the results more stable.  As we mentioned, because $w$ combines signals in different bins at the level of likelihood while $ \xip $ works at the level of data vector, this also contributes to the stability of $w$ relative to $\xip$. This is apparent in the missing bin tests.

\change{ DES Y6 shares the same footprint as DES Y3 but with deeper magnitude, and so the BAO sample will have higher number density and will extend to redshift 1.2. It will undergo a new round of photo-$z$ and other systematics check, and these will help further verify if there is untreated systematics contaminations, especially in the high redshift bins.   Whether $\xip$ will give strong constraint in this case will depend on the final dataset, but in any case, it offers an important means to crosscheck with the conventional angular statistics results.   }

\section{Conclusions}
\label{sec:conslusions}

We have presented  measurements of the BAO scale using the 3D correlation projected to the transverse scale, $\xip$. The dataset is derived from the DES Y3 and the final sample consists of about 7 million galaxies in the redshift range of [0.6,1.1] over a footprint of 4108 deg$^2$. Although the  angular correlation function and angular power spectrum had been applied to this sample to measure the BAO in DES Y3  \cite{DES:2021esc}, the treatments in these statistics are similar in many aspects. On the other hand, $\xip$ is less correlated with the angular statistics and can serve as an independent check.  We have systematically compared the  $\xip$ results against those from the angular correlation function $w$, which has been a benchmark for the tomographic analysis of the photometric data.

Our work follows the improved modeling in \cite{Chan_etal2021}.  In particular, realistic photo-$z$ distribution is incorporated in the template modeling and the Gaussian covariance computation. These overcome the shortcomings of the Gaussian photo-$z$ approximation in previous works and help us to isolate the remaining potential systematics in the $\xip$ method.    The $\xip$ statistic is obtained by averaging over $\mu$ pairs with a suitable stacking window.  We have presented results for two windows. The first one is a top-hat [Eq.~\eqref{eq:TH_stacking}] with $ \mu_{\rm max}=0.8 $ \cite{Ross:2017emc}, which had always been assumed in previous works. This is the blinded result, which has passed a battery of robustness tests similar to those of $w$ and $C_\ell $ in DES Y3.    However, we point out that the signal-to-noise decreases as $\mu$ increases, equal weighting is sub-optimal. We propose a cut-off Gaussian window [Eq.~\eqref{eq:cutoffGaussian_stacking}], which downweights the high $\mu $ pairs in favor of the low $\mu$ ones. This window increases the stability of the $\xip$ method, and we have verified that with the mock test results.   Although the Gaussian window is adopted after unblinding, the updated pipeline does not bias the results because the best fit is not sensitive to the precise value of  $\sigma_\mu $ and the choice of $\sigma_\mu =0.3 $ reflects the average error bar size  in the low  $\sigma_\mu $ vicinity.

For the full sample, we have measured  $ D_{\rm M} / r_{\rm s} $ to be $19.15 \pm 0.58  $ for $  W_{\rm G} $ (unblinded)  and  $ 19.00 \pm  0.67 $ for  $ W_{\rm TH}  $ (blinded).  Especially for $ W_{\rm TH} $, the resultant error bar is bigger than that from the angular correlation function $w$, $ 18.84 \pm  0.50 $. 
The deviation from Planck results is reduced to 1.6 $\sigma $ (1.7 $\sigma$)  for $  W_{\rm G} $ ($W_{\rm TH} $) and it is less significant than DES Y3 $w$ analysis.   On the other hand, we find that for individual redshift bin fit, $  W_{\rm G} $  actually gives a tighter bound than $w$. We deduce that the poor error bound for the whole sample is caused by the BAO signals in the full sample being heterogeneous and hence the total BAO signal is smeared out. From the mock test, we also find that if the sample is heterogeneous in BAO signals, the chance that  $\xip$ yields a larger error bar than $w$ is enhanced.   We then consider a sub-sample with more consistent BAO signals composed of data in the redshift range  $ 0.7< z_{\rm p} < 1 $. The effective redshift for the homogeneous sample is 0.845.  The constraint on  $ D_{\rm M} / r_{\rm s} $ is  $19.84 \pm 0.53  $ for $  W_{\rm G} $ and  $ 19.80 \pm  0.67 $ for  $ W_{\rm TH} $. For  $  W_{\rm G} $, the error bound is tighter than the corresponding $w$ result, which reads $ 19.86 \pm 0.55 $.

We conducted numerous robustness tests to check the stability of the results.  Overall we find that  $w$ yields the most stable results, $  W_{\rm G} $ the second, and $  W_{\rm TH} $ the least stable. These tests also help us better understand the properties of these statistics.   First, they differ in the order of projection and correlation measurement. Because for $w$, the data is projected to the angular space first, the effects of the photo-$z$ contamination can be effectively limited.  By measuring the 3D correlation first,  $\xip$ not only measures the transverse information but also some radial signals, but this also allows the photo-$z$ noise to sneak in and cause some instability in the results.   Second, they treat the signals in the tomographic bins differently. For $ \xip $, the signals in the whole dataset are combined primitively, at the level of data vector, while for $w$, they are combined at the level of likelihood.  This causes $w$ to be more robust to heterogeneous signals, and the contribution of signals from individual bins to be more predictable.  For heterogeneous signals, $w$ can give a tight error bound, while $ \xip $ tends to give a loose bound.  Thus it is important to verify that the heterogeneous signals are self-consistent, otherwise $w$ can give an artificially tight bound. Conversely, $ \xip $ gives a loose bound for heterogeneous signals does not necessarily imply that it is an inferior statistic because its weak bound can reflect potential systematics in the data or hints of deviation of the underlying cosmological model.

Our analysis further clarifies the properties of the $\xip$ method and demonstrates its utilities.  $\xip$ and $w$ (or angular statistics in general) have their own advantages and drawbacks, and they can crosscheck each other.   We anticipate that $\xip$ will continue to play an important role in the  forthcoming imaging data analysis such as DES Y6 and other photometric surveys mentioned in the introduction.

\section*{Acknowledgments}
K.C.C. acknowledges the support from the National Science Foundation of China under the grant 11873102 and 12273121, the science research grants from the China Manned Space Project with NO.CMS-CSST-2021-B01, and the Science and Technology Program of Guangzhou, China (No. 202002030360).   S.A. is supported by the Spanish Agencia Estatal de Investigacion through the grant “IFT Centro de Excelencia Severo Ochoa by CEX2020-001007-S" and ``EU-HORIZON-2020-776247 Enabling Weak Lensing Cosmology (EWC)'', and was also supported by Atraccion de Talento program no. 2019-T1/TIC-12702 granted by the Comunidad de Madrid in Spain.

Funding for the DES Projects has been provided by the U.S. Department of Energy, the U.S. National Science Foundation, the Ministry of Science and Education of Spain, 
the Science and Technology Facilities Council of the United Kingdom, the Higher Education Funding Council for England, the National Center for Supercomputing 
Applications at the University of Illinois at Urbana-Champaign, the Kavli Institute of Cosmological Physics at the University of Chicago, 
the Center for Cosmology and Astro-Particle Physics at the Ohio State University,
the Mitchell Institute for Fundamental Physics and Astronomy at Texas A\&M University, Financiadora de Estudos e Projetos, 
Funda{\c c}{\~a}o Carlos Chagas Filho de Amparo {\`a} Pesquisa do Estado do Rio de Janeiro, Conselho Nacional de Desenvolvimento Cient{\'i}fico e Tecnol{\'o}gico and 
the Minist{\'e}rio da Ci{\^e}ncia, Tecnologia e Inova{\c c}{\~a}o, the Deutsche Forschungsgemeinschaft and the Collaborating Institutions in the Dark Energy Survey. 

The Collaborating Institutions are Argonne National Laboratory, the University of California at Santa Cruz, the University of Cambridge, Centro de Investigaciones Energ{\'e}ticas, 
Medioambientales y Tecnol{\'o}gicas-Madrid, the University of Chicago, University College London, the DES-Brazil Consortium, the University of Edinburgh, 
the Eidgen{\"o}ssische Technische Hochschule (ETH) Z{\"u}rich, 
Fermi National Accelerator Laboratory, the University of Illinois at Urbana-Champaign, the Institut de Ci{\`e}ncies de l'Espai (IEEC/CSIC), 
the Institut de F{\'i}sica d'Altes Energies, Lawrence Berkeley National Laboratory, the Ludwig-Maximilians Universit{\"a}t M{\"u}nchen and the associated Excellence Cluster Universe, 
the University of Michigan, NSF's NOIRLab, the University of Nottingham, The Ohio State University, the University of Pennsylvania, the University of Portsmouth, 
SLAC National Accelerator Laboratory, Stanford University, the University of Sussex, Texas A\&M University, and the OzDES Membership Consortium.

Based in part on observations at Cerro Tololo Inter-American Observatory at NSF's NOIRLab (NOIRLab Prop. ID 2012B-0001; PI: J. Frieman), which is managed by the Association of Universities for Research in Astronomy (AURA) under a cooperative agreement with the National Science Foundation.

The DES data management system is supported by the National Science Foundation under Grant Numbers AST-1138766 and AST-1536171.
The DES participants from Spanish institutions are partially supported by MICINN under grants ESP2017-89838, PGC2018-094773, PGC2018-102021, SEV-2016-0588, SEV-2016-0597, and MDM-2015-0509, some of which include ERDF funds from the European Union. IFAE is partially funded by the CERCA program of the Generalitat de Catalunya.
Research leading to these results has received funding from the European Research
Council under the European Union's Seventh Framework Program (FP7/2007-2013) including ERC grant agreements 240672, 291329, and 306478.
We  acknowledge support from the Brazilian Instituto Nacional de Ci\^encia
e Tecnologia (INCT) do e-Universo (CNPq grant 465376/2014-2).

This manuscript has been authored by Fermi Research Alliance, LLC under Contract No. DE-AC02-07CH11359 with the U.S. Department of Energy, Office of Science, Office of High Energy Physics.

\appendix

\section{Pre-unblinding test} 
\label{sec:PreunblindingTest}

To avoid the possibility of confirmation bias, in \cite{DES:2021esc}, the cosmologically interesting part of the data is blinded from the analysts until the pipeline is finalized.    A set of pre-unblinding tests were performed to check the validility of  the methodology and the data in particular.  Only after the tests are passed, the data are considered ready for cosmological analysis and hence unblinded.  For this work, although the BAO sample had been unblinded, we follow the DES practice to carry out the pre-unblinding tests laid down in \cite{DES:2021esc}.  They serve as additional tests to the $\xip$ statistic.  Many of  these tests are similar to the robustness tests presented in Sec.~\ref{sec:robustness_test}.  Note that the Gaussian window is adopted after unblinding, and so these tests only incudes the $W_{\rm TH} $  results.  Nonetheless, for completeness, we  show the pre-unblinding results here for reference.


The results are shown in Table \ref{tab:preunblind_test}. These tests check how the best fit value and the estimated error bar change in response to  removing one of the tomographic bins, using the alternative Planck template, Planck covariance or the \ZMC photo-$z$ estimation. The confidence intervals are derived by applying the tests to the mock catalog.   The tests on the mocks are performed in MICE fiducial cosmology,  and for the data, the tests are done in both MICE and Planck cosmologies.   If all the test results  on the data fall within the 90\% intervals, the tests are considered passed.   If some of the test results only satisfy more extreme intervals, then further conditions  are used to judge the ``normality'' of the results. We find that  all the test results fall within the 90\% interval, and hence the tests are passed.

Note that the mocks are designed to follow the VIPERS distribution, but there is no exact \ZMC analog on the mocks.   To this end, we directly use the \ZMC distribution for the data in the mock test. We find that the intervals are negatively biased.  To investigate this further we perform a test using the original VIPERS distribution, and the results are aslo shown in Table \ref{tab:preunblind_test}. The intervals are also found to be shifted to the negative side slightly. We conclude that the shift to the negative side could be caused by  the photo-$z$ distribution calibation in the mock construction.

\begin{table*}[]
  \caption{
    The pre-unblinding tests for  $\xip$ with $W_{\rm TH} $, showing the impact of removing data in individual tomographic bins, of changing the assumed cosmology for the BAO template or the covariance, and of considering an independent estimate of the true redshift distributions. Both the change in the best fit values and the error bars are shown. The confidence intervals are derived from the mock. The mock tests are done in MICE cosmology, but both MICE and Planck fiducial cosmologies are considered in the actual data tests. All the data results fall within the 90\% interval, and so the pre-unblinding tests are considered passed.         \\}
\label{tab:preunblind_test}
\begin{tabular}{l|ll|ll|ll|ll||l|l}
\hline
\hline
Threshold       & \multicolumn{2}{l|}{0.9} & \multicolumn{2}{l|}{0.95} & \multicolumn{2}{l|}{0.97} & \multicolumn{2}{l||}{0.99} & \multicolumn{2}{l}{data} \\ \cline{2-11}
(Fraction of mocks)   &   min &          max   &          min &          max &          min&          max &          min&          max & {\tt MICE}       & {\tt Planck}       \\ \hline
                \T &   \multicolumn{8}{c||}{$10^2 (\alpha-\alpha_{\rm fiducial})$}  \B                                                                                         \\ \cline{2-11}            
No bin 1                      &  -3.13      & 2.68      & -3.55      & 3.61        & -4.16       & 4.75       & -4.49       & 5.62              & 2.42     & 1.49     \\
No bin 2                      & -2.72       & 2.62      & -3.04       & 3.21       & -4.26       & 3.29       & -5.56       & 3.84              & -0.82    &-1.48     \\
No bin 3                      & -2.73       & 2.40      & -2.97       & 2.88       & -3.21       & 4.24       & -3.43       & 5.54              & -0.74    &-0.44       \\
No bin 4                      & -1.52       & 2.54      & -1.71       & 3.12       & -2.19       & 3.35       & -2.90       & 3.53              & 0.10     &-0.24     \\
No bin 5                      & -1.19       & 1.80      & -1.55       & 2.11       & -1.61       & 2.36       & -1.85       & 2.88              & 0.66     & 1.37     \\
{\tt Planck} Template         & -2.66       & 1.56      & -2.97       & 1.61       & -3.46       & 2.18       & -4.62       & 2.62              & 0.34     & ---           \\
{\tt Planck} Covariance       & -0.67       & 0.56      & -1.01       & 0.59       & -1.06       & 0.62       & -1.27       & 0.68              &-0.42     & ---           \\
$n(z)_{{\rm DNF} - { Z_{\rm MC} } }$  & -1.13       & -0.20     & -1.20       &-0.15       & -1.24    &-0.01     & -1.34       &0.08                    & -0.18    & -0.24         \\
\hline
original VIPERS              & -1.24       & 0.42     & -1.31       &0.92       & -1.34    & 1.03     & -1.57       &  1.16                      & --    &  --       \\

\hline
            \T    & \multicolumn{8}{c||}{$(\sigma-\sigma_{\rm All \ Bins})/\sigma_{\rm All \ Bins}$}  \B                                                                        \\ \cline{2-11}            
No bin 1         & -0.24               & 0.45        & -0.30     & 0.49        & -0.30       & 0.53       & -0.40      & 0.70                & -0.03      & -0.08        \\
No bin 2         & -0.22               & 0.50        & -0.24     & 0.59        & -0.25       & 0.73       & -0.26      & 0.86                & 0.37       & 0.02         \\
No bin 3         & -0.11               & 0.77        & -0.13     & 0.87        & -0.20       & 0.94       & -0.25      & 1.22                & 0.16      & 0.13       \\
No bin 4         & -0.12               & 0.35        & -0.17     & 0.38        & -0.20       & 0.40       & -0.30      & 0.61                & 0.35       & 0.01       \\
No bin 5         & -0.16               & 0.24        & -0.22     & 0.32        & -0.23       & 0.34       & -0.35      & 0.38                & 0.06      & -0.14         \\
\hline
\end{tabular}
\end{table*}

\section{ Error bars in the mocks with heterogeneous signals }

In the main text, we find that the error bar derived from $\xip$  is larger than that obtained from $w$ by a significant fraction. On the other hand, when the fits are performed on the individual tomographic bins or the homogeneous BAO-signal sample bins 2, 3, and 4, the errors obtained from $\xip$ is tighter or compatible with that from $w$.  We conclude that the error bar from $\xip$ is loosened when the signals are heterogeneous, while the angular correlation function seems less affected. In this appendix, we shall further investigate the impact of the heterogeneity using the mocks.


We first perform the BAO fit on the individual tomographic bins each of width $ \Delta z = 0.1$ in the redshift range [0.6,1.1].  Owing to limitation in computing power, we only use 93 mocks (one mock is removed because no detection for $\xip$  in one of the bins).  We use the standard deviation of the best fit to the individual tomographic bins to select mocks with heterogeneous BAO signals.  We illustrate this with the Gaussian window with  $ \sigma_\mu = 0.3 $.  The average of the standard deviation of the best fit to the individual tomographic bins is 0.049 for $\xip$ (0.051 for $w$).  Because there is no detection for the first bin, to compute the standard deviation for the actual data fit, we assign the $\alpha$ value to be the lower boundary, 0.8.  Both $\xip$ and  $w$ give the same   standard deviation, 0.07.  Thus, we take the mocks with the standard deviation value larger than 0.07 as the heterogeneous mocks.  This threshold also ensures that there are decent number of mocks available.  The set of mocks satisfying this heterogeneity condition differs slightly for $\xip$ and $w$.  We then end up with 23 mocks for $\xip$ (22 for $w$).

\begin{figure}[!htb]
\centering
\includegraphics[width=\linewidth]{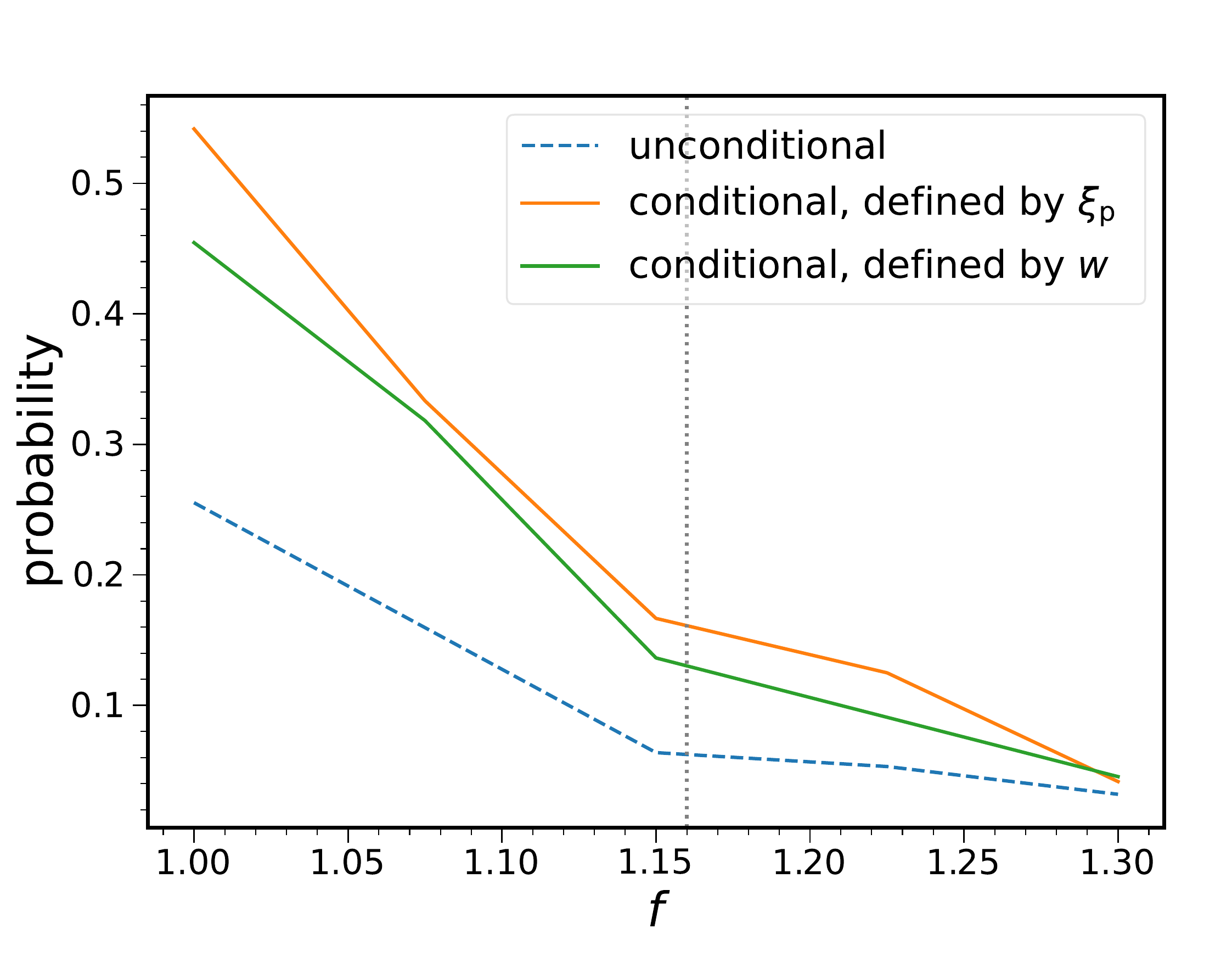}
\caption{  The probability of finding mocks with error bars derived from $\xip$ being larger than that from $w$ by at least a factor of $f$. The results for all the mocks (unconditional, dashed blue) and conditioned on the heterogeneous  BAO-signal  mocks (solid). Both the results for the heterogeneous mocks defined by $\xip$ (orange) and $w$ (green) are shown.  Here the results are obtained with  $W_{\rm G} $.   The probability is increased when conditioned on the heterogeneous mocks. The $f$ value corresponding to the actual data fit is indicated by a dotted vertical line.   }  
  \label{fig:err_heterog_mocks}
\end{figure}

We consider the probability for the condition $\sigma_{\xip} > f \sigma_w $, where $\sigma_{\xip}$ and $\sigma_w $ denote the error bars derived from the BAO fit using $\xip$ and $w$ on the data including all five bins and $ f $ is a parameter. In Fig.~\ref{fig:err_heterog_mocks},  we plot the probability as a function of $f$.

We show the unconditional case computed with all the available mocks, and the conditional case  obtained with the heterogeneous BAO-signal mocks. For the conditional cases, we show  both the results derived from the heterogeneous mocks defined with  $\xip$ and $w$, respectively. We indeed find that there is an increased probability of finding mocks meeting the condition $\sigma_{\xip} > f \sigma_w $ compared to the unconditional one.  We caution that  because the size of the heterogeneous sample is small, the probability obtained only serves as a rough estimation. This is even more true for the probability in the high $f$ end.  We also indicates the $f$ value corresponding to the actual data results on the plot, and our estimate gives a probability of 16\% for this heterogeneous sample.

For $ W_{\rm TH} $, we also find that the probability is enhanced relative to the unconditional case. Unlike the  $ W_{\rm G} $ case, it is enhanced even at $f=1.3$, and this trend agrees with the fact that  $ W_{\rm TH} $ yields an even larger error bar in the data fit.   Furthermore, we have considered alternative definition of heterogeneous mock by utilizing  the difference between the maximum and the minimum among the tomographic bin fit results, and the results are qualitatively similar.

\label{appendix:errorbars}

\bibliography{references}

\end{document}

%% file: author_list.tex


\author{K.~C.~Chan}
\email{chankc@mail.sysu.edu.cn}
\affiliation{ School of Physics and Astronomy, Sun Yat-sen University, 2 Daxue Road, Tangjia, Zhuhai 519082, China}
\affiliation{CSST Science Center for the Guangdong-Hongkong-Macau Greater Bay Area, SYSU, Zhuhai 519082,China}
\author{S.~Avila}
\affiliation{Instituto de Fisica Teorica UAM/CSIC, Universidad Autonoma de Madrid, 28049 Madrid, Spain}
\affiliation{Departamento de F\'isica Te\'orica,  Universidad Aut\'onoma de Madrid, 28049 Madrid, Spain}
\author{A.~Carnero~Rosell}
\affiliation{Instituto de Astrofisica de Canarias, E-38205 La Laguna, Tenerife, Spain}
\affiliation{Laborat\'orio Interinstitucional de e-Astronomia - LIneA, Rua Gal. Jos\'e Cristino 77, Rio de Janeiro, RJ - 20921-400, Brazil}
\affiliation{Universidad de La Laguna, Dpto. Astrofísica, E-38206 La Laguna, Tenerife, Spain}
\author{I.~Ferrero}
\affiliation{Institute of Theoretical Astrophysics, University of Oslo. P.O. Box 1029 Blindern, NO-0315 Oslo, Norway}
\author{J.~Elvin-Poole}
\affiliation{Center for Cosmology and Astro-Particle Physics, The Ohio State University, Columbus, OH 43210, USA}
\affiliation{Department of Physics, The Ohio State University, Columbus, OH 43210, USA}
\author{E.~Sanchez}
\affiliation{Centro de Investigaciones Energ\'eticas, Medioambientales y Tecnol\'ogicas (CIEMAT), Madrid, Spain}
\author{H.~Camacho}
\affiliation{Instituto de F\'{i}sica Te\'orica, Universidade Estadual Paulista, S\~ao Paulo, Brazil}
\affiliation{Laborat\'orio Interinstitucional de e-Astronomia - LIneA, Rua Gal. Jos\'e Cristino 77, Rio de Janeiro, RJ - 20921-400, Brazil}
\author{A.~Porredon}
\affiliation{Center for Cosmology and Astro-Particle Physics, The Ohio State University, Columbus, OH 43210, USA}
\affiliation{Department of Physics, The Ohio State University, Columbus, OH 43210, USA}
\affiliation{Institute for Astronomy, University of Edinburgh, Edinburgh EH9 3HJ, UK}
\author{M.~Crocce}
\affiliation{Institut d'Estudis Espacials de Catalunya (IEEC), 08034 Barcelona, Spain}
\affiliation{Institute of Space Sciences (ICE, CSIC),  Campus UAB, Carrer de Can Magrans, s/n,  08193 Barcelona, Spain}
\author{T.~M.~C.~Abbott}
\affiliation{Cerro Tololo Inter-American Observatory, NSF's National Optical-Infrared Astronomy Research Laboratory, Casilla 603, La Serena, Chile}
\author{M.~Aguena}
\affiliation{Laborat\'orio Interinstitucional de e-Astronomia - LIneA, Rua Gal. Jos\'e Cristino 77, Rio de Janeiro, RJ - 20921-400, Brazil}
\author{S.~Allam}
\affiliation{Fermi National Accelerator Laboratory, P. O. Box 500, Batavia, IL 60510, USA}
\author{F.~Andrade-Oliveira}
\affiliation{Department of Physics, University of Michigan, Ann Arbor, MI 48109, USA}
\author{E.~Bertin}
\affiliation{CNRS, UMR 7095, Institut d'Astrophysique de Paris, F-75014, Paris, France}
\affiliation{Sorbonne Universit\'es, UPMC Univ Paris 06, UMR 7095, Institut d'Astrophysique de Paris, F-75014, Paris, France}
\author{S.~Bocquet}
\affiliation{University Observatory, Faculty of Physics, Ludwig-Maximilians-Universit\"at, Scheinerstr. 1, 81679 Munich, Germany}
\author{D.~Brooks}
\affiliation{Department of Physics \& Astronomy, University College London, Gower Street, London, WC1E 6BT, UK}
\author{D.~L.~Burke}
\affiliation{Kavli Institute for Particle Astrophysics \& Cosmology, P. O. Box 2450, Stanford University, Stanford, CA 94305, USA}
\affiliation{SLAC National Accelerator Laboratory, Menlo Park, CA 94025, USA}
\author{M.~Carrasco~Kind}
\affiliation{Center for Astrophysical Surveys, National Center for Supercomputing Applications, 1205 West Clark St., Urbana, IL 61801, USA}
\affiliation{Department of Astronomy, University of Illinois at Urbana-Champaign, 1002 W. Green Street, Urbana, IL 61801, USA}
\author{J.~Carretero}
\affiliation{Institut de F\'{\i}sica d'Altes Energies (IFAE), The Barcelona Institute of Science and Technology, Campus UAB, 08193 Bellaterra (Barcelona) Spain}
\author{F.~J.~Castander}
\affiliation{Institut d'Estudis Espacials de Catalunya (IEEC), 08034 Barcelona, Spain}
\affiliation{Institute of Space Sciences (ICE, CSIC),  Campus UAB, Carrer de Can Magrans, s/n,  08193 Barcelona, Spain}
\author{R.~Cawthon}
\affiliation{Physics Department, William Jewell College, Liberty, MO, 64068}
\author{C.~Conselice}
\affiliation{Jodrell Bank Center for Astrophysics, School of Physics and Astronomy, University of Manchester, Oxford Road, Manchester, M13 9PL, UK}
\affiliation{University of Nottingham, School of Physics and Astronomy, Nottingham NG7 2RD, UK}
\author{M.~Costanzi}
\affiliation{Astronomy Unit, Department of Physics, University of Trieste, via Tiepolo 11, I-34131 Trieste, Italy}
\affiliation{INAF-Osservatorio Astronomico di Trieste, via G. B. Tiepolo 11, I-34143 Trieste, Italy}
\affiliation{Institute for Fundamental Physics of the Universe, Via Beirut 2, 34014 Trieste, Italy}
\author{M.~E.~S.~Pereira}
\affiliation{Hamburger Sternwarte, Universit\"{a}t Hamburg, Gojenbergsweg 112, 21029 Hamburg, Germany}
\author{J.~De~Vicente}
\affiliation{Centro de Investigaciones Energ\'eticas, Medioambientales y Tecnol\'ogicas (CIEMAT), Madrid, Spain}
\author{S.~Desai}
\affiliation{Department of Physics, IIT Hyderabad, Kandi, Telangana 502285, India}
\author{H.~T.~Diehl}
\affiliation{Fermi National Accelerator Laboratory, P. O. Box 500, Batavia, IL 60510, USA}
\author{P.~Doel}
\affiliation{Department of Physics \& Astronomy, University College London, Gower Street, London, WC1E 6BT, UK}
\author{S.~Everett}
\affiliation{Jet Propulsion Laboratory, California Institute of Technology, 4800 Oak Grove Dr., Pasadena, CA 91109, USA}
\author{B.~Flaugher}
\affiliation{Fermi National Accelerator Laboratory, P. O. Box 500, Batavia, IL 60510, USA}
\author{P.~Fosalba}
\affiliation{Institut d'Estudis Espacials de Catalunya (IEEC), 08034 Barcelona, Spain}
\affiliation{Institute of Space Sciences (ICE, CSIC),  Campus UAB, Carrer de Can Magrans, s/n,  08193 Barcelona, Spain}
\author{J.~Garc\'ia-Bellido}
\affiliation{Instituto de Fisica Teorica UAM/CSIC, Universidad Autonoma de Madrid, 28049 Madrid, Spain}
\author{E.~Gaztanaga}
\affiliation{Institut d'Estudis Espacials de Catalunya (IEEC), 08034 Barcelona, Spain}
\affiliation{Institute of Space Sciences (ICE, CSIC),  Campus UAB, Carrer de Can Magrans, s/n,  08193 Barcelona, Spain}
\author{D.~W.~Gerdes}
\affiliation{Department of Astronomy, University of Michigan, Ann Arbor, MI 48109, USA}
\affiliation{Department of Physics, University of Michigan, Ann Arbor, MI 48109, USA}
\author{T.~Giannantonio}
\affiliation{Institute of Astronomy, University of Cambridge, Madingley Road, Cambridge CB3 0HA, UK}
\affiliation{Kavli Institute for Cosmology, University of Cambridge, Madingley Road, Cambridge CB3 0HA, UK}
\author{D.~Gruen}
\affiliation{University Observatory, Faculty of Physics, Ludwig-Maximilians-Universit\"at, Scheinerstr. 1, 81679 Munich, Germany}
\author{R.~A.~Gruendl}
\affiliation{Center for Astrophysical Surveys, National Center for Supercomputing Applications, 1205 West Clark St., Urbana, IL 61801, USA}
\affiliation{Department of Astronomy, University of Illinois at Urbana-Champaign, 1002 W. Green Street, Urbana, IL 61801, USA}
\author{G.~Gutierrez}
\affiliation{Fermi National Accelerator Laboratory, P. O. Box 500, Batavia, IL 60510, USA}
\author{S.~R.~Hinton}
\affiliation{School of Mathematics and Physics, University of Queensland,  Brisbane, QLD 4072, Australia}
\author{D.~L.~Hollowood}
\affiliation{Santa Cruz Institute for Particle Physics, Santa Cruz, CA 95064, USA}
\author{K.~Honscheid}
\affiliation{Center for Cosmology and Astro-Particle Physics, The Ohio State University, Columbus, OH 43210, USA}
\affiliation{Department of Physics, The Ohio State University, Columbus, OH 43210, USA}
\author{D.~Huterer}
\affiliation{Department of Physics, University of Michigan, Ann Arbor, MI 48109, USA}
\author{D.~J.~James}
\affiliation{Center for Astrophysics $\vert$ Harvard \& Smithsonian, 60 Garden Street, Cambridge, MA 02138, USA}
\author{K.~Kuehn}
\affiliation{Australian Astronomical Optics, Macquarie University, North Ryde, NSW 2113, Australia}
\affiliation{Lowell Observatory, 1400 Mars Hill Rd, Flagstaff, AZ 86001, USA}
\author{O.~Lahav}
\affiliation{Department of Physics \& Astronomy, University College London, Gower Street, London, WC1E 6BT, UK}
\author{C.~Lidman}
\affiliation{Centre for Gravitational Astrophysics, College of Science, The Australian National University, ACT 2601, Australia}
\affiliation{The Research School of Astronomy and Astrophysics, Australian National University, ACT 2601, Australia}
\author{M.~Lima}
\affiliation{Departamento de F\'isica Matem\'atica, Instituto de F\'isica, Universidade de S\~ao Paulo, CP 66318, S\~ao Paulo, SP, 05314-970, Brazil}
\affiliation{Laborat\'orio Interinstitucional de e-Astronomia - LIneA, Rua Gal. Jos\'e Cristino 77, Rio de Janeiro, RJ - 20921-400, Brazil}
\author{J.~L.~Marshall}
\affiliation{George P. and Cynthia Woods Mitchell Institute for Fundamental Physics and Astronomy, and Department of Physics and Astronomy, Texas A\&M University, College Station, TX 77843,  USA}
\author{J. Mena-Fern{\'a}ndez}
\affiliation{Centro de Investigaciones Energ\'eticas, Medioambientales y Tecnol\'ogicas (CIEMAT), Madrid, Spain}
\author{F.~Menanteau}
\affiliation{Center for Astrophysical Surveys, National Center for Supercomputing Applications, 1205 West Clark St., Urbana, IL 61801, USA}
\affiliation{Department of Astronomy, University of Illinois at Urbana-Champaign, 1002 W. Green Street, Urbana, IL 61801, USA}
\author{R.~Miquel}
\affiliation{Instituci\'o Catalana de Recerca i Estudis Avan\c{c}ats, E-08010 Barcelona, Spain}
\affiliation{Institut de F\'{\i}sica d'Altes Energies (IFAE), The Barcelona Institute of Science and Technology, Campus UAB, 08193 Bellaterra (Barcelona) Spain}
\author{A.~Palmese}
\affiliation{Department of Astronomy, University of California, Berkeley,  501 Campbell Hall, Berkeley, CA 94720, USA}
\author{F.~Paz-Chinch\'{o}n}
\affiliation{Center for Astrophysical Surveys, National Center for Supercomputing Applications, 1205 West Clark St., Urbana, IL 61801, USA}
\affiliation{Institute of Astronomy, University of Cambridge, Madingley Road, Cambridge CB3 0HA, UK}
\author{A.~Pieres}
\affiliation{Laborat\'orio Interinstitucional de e-Astronomia - LIneA, Rua Gal. Jos\'e Cristino 77, Rio de Janeiro, RJ - 20921-400, Brazil}
\affiliation{Observat\'orio Nacional, Rua Gal. Jos\'e Cristino 77, Rio de Janeiro, RJ - 20921-400, Brazil}
\author{A.~A.~Plazas~Malag\'on}
\affiliation{Department of Astrophysical Sciences, Princeton University, Peyton Hall, Princeton, NJ 08544, USA}
\author{M.~Raveri}
\affiliation{Department of Physics and Astronomy, University of Pennsylvania, Philadelphia, PA 19104, USA}
\author{M.~Rodriguez-Monroy}
\affiliation{Centro de Investigaciones Energ\'eticas, Medioambientales y Tecnol\'ogicas (CIEMAT), Madrid, Spain}
\author{A.~Roodman}
\affiliation{Kavli Institute for Particle Astrophysics \& Cosmology, P. O. Box 2450, Stanford University, Stanford, CA 94305, USA}
\affiliation{SLAC National Accelerator Laboratory, Menlo Park, CA 94025, USA}
\author{A.~J.~Ross}
\affiliation{Center for Cosmology and Astro-Particle Physics, The Ohio State University, Columbus, OH 43210, USA}
\author{V.~Scarpine}
\affiliation{Fermi National Accelerator Laboratory, P. O. Box 500, Batavia, IL 60510, USA}
\author{I.~Sevilla-Noarbe}
\affiliation{Centro de Investigaciones Energ\'eticas, Medioambientales y Tecnol\'ogicas (CIEMAT), Madrid, Spain}
\author{M.~Smith}
\affiliation{School of Physics and Astronomy, University of Southampton,  Southampton, SO17 1BJ, UK}
\author{E.~Suchyta}
\affiliation{Computer Science and Mathematics Division, Oak Ridge National Laboratory, Oak Ridge, TN 37831}

\author{M.~E.~C.~Swanson}
\affiliation{Center for Astrophysical Surveys, National Center for Supercomputing Applications, 1205 West Clark St., Urbana, IL 61801, USA}

\author{G.~Tarle}
\affiliation{Department of Physics, University of Michigan, Ann Arbor, MI 48109, USA}
\author{D.~Thomas}
\affiliation{Institute of Cosmology and Gravitation, University of Portsmouth, Portsmouth, PO1 3FX, UK}
\author{D.~L.~Tucker}
\affiliation{Fermi National Accelerator Laboratory, P. O. Box 500, Batavia, IL 60510, USA}
\author{M.~Vincenzi}
\affiliation{Institute of Cosmology and Gravitation, University of Portsmouth, Portsmouth, PO1 3FX, UK}
\affiliation{School of Physics and Astronomy, University of Southampton,  Southampton, SO17 1BJ, UK}
\author{N.~Weaverdyck}
\affiliation{Department of Physics, University of Michigan, Ann Arbor, MI 48109, USA}
\affiliation{Lawrence Berkeley National Laboratory, 1 Cyclotron Road, Berkeley, CA 94720, USA}

\collaboration{DES Collaboration}
